\documentclass[journal]{IEEEtran}
\usepackage{multirow}
\usepackage{booktabs}
\usepackage{amsfonts}
\usepackage{amsmath}
\usepackage{amsmath,cases}
\usepackage{amssymb}
\usepackage{graphicx}
\usepackage{cite}
\usepackage{epsfig}
\usepackage{url}
\usepackage{algorithm}
\usepackage{algorithmicx, algpseudocode}
\usepackage{epstopdf}
\usepackage{color}
\usepackage[tight]{subfigure}
\usepackage{mathrsfs}
\usepackage[justification=centering]{caption}
\usepackage{float}
\usepackage{diagbox}
\usepackage{balance}

\newcommand{\ds}{\displaystyle}
\newtheorem{myth}{Theorem}

\newcommand{\la}{\langle}
\newcommand{\ra}{\rangle}
\newcommand{\bphi}{\pmb{\phi}}

\newcommand{\bx}{\boldsymbol{x}}
\newcommand{\qed}{\mbox{}\hfill$\Box$\vskip 3mm}
\newcommand{\Prf}{{\it Proof: \ }}

\newcommand{\clL}{{\cal L}}
\newcommand{\clN}{{\cal N}}

\newcommand{\clK}{{\cal K}}

\newcommand{\clI}{{\cal I}}

\newcommand{\bv}{\mathbf{v}}

\newcommand{\clB}{{\cal B}}

\newcommand{\by}{\mathbf{y}}

\newcommand{\bfy}{\mathbf{y}}
\newcommand{\xk}{x^{(\kappa)}}
\newcommand{\zk}{z^{(\kappa)}}
\newcommand{\alphak}{\alpha^{(\kappa)}}
\newcommand{\talphak}{\tilde{\alpha}^{(\kappa)}}

\newcommand{\rk}{r^{(\kappa)}}
\newcommand{\ak}{a^{(\kappa)}}
\newcommand{\dk}{d^{(\kappa)}}
\newcommand{\tdk}{\tilde{d}^{(\kappa)}}
\newcommand{\ck}{c^{(\kappa)}}
\newcommand{\tck}{\tilde{c}^{(\kappa)}}
\newcommand{\tbk}{\tilde{b}^{(\kappa)}}

\newcommand{\Ck}{C^{(\kappa)}}

\newcommand{\rhok}{\rho^{(\kappa)}}

\newcommand{\bk}{b^{(\kappa)}}

\newcommand{\trk}{\tilde{r}^{(\kappa)}}

\newcommand{\btheta}{\pmb{\theta}}

\newcommand{\fk}{f^{(\kappa)}}

\newcommand{\thetak}{\theta^{(\kappa)}}
\newcommand{\betak}{\beta^{(\kappa)}}
\newcommand{\tbetak}{\tilde{\beta}^{(\kappa)}}

\newcommand{\varphik}{\varphi^{(\kappa)}}

\newcommand{\thetako}{\theta^{(\kappa+1)}}
\newcommand{\Phik}{\Phi^{(\kappa)}}
\newcommand{\tak}{\tilde{a}^{(\kappa)}}

\newcommand{\vBk}{v^{B,(\kappa)}}
\newcommand{\vBko}{v^{B, (\kappa+1)}}

\newcommand{\tAk}{\tilde{A}^{(\kappa)}}

\newcommand{\trhok}{\tilde{\rho}^{(\kappa)}}

\newcommand{\tCk}{\tilde{C}^{(\kappa)}}

\newcommand{\bPhi}{\pmb{\Phi}}
\newcommand{\Phiko}{\Phi^{(\kappa+1)}}
\newcommand{\hko}{h^{(\kappa+1)}}

\newcommand{\clAko}{\mathcal{A}^{(\kappa+1)}}

\newcommand{\tih}{\tilde{h}}
\newcommand{\phik}{\phi^{(\kappa)}}
\newcommand{\phiko}{\phi^{(\kappa+1)}}
\newcommand{\Xik}{\Xi^{(\kappa)}}
\newcommand{\tXik}{\tilde{\Xi}^{(\kappa)}}
\newcommand{\balphak}{\bar{\alpha}^{(\kappa)}}
\newcommand{\yk}{y^{(\kappa)}}
\newcommand{\txk}{\tilde{x}^{(\kappa)}}
\newcommand{\tyk}{\tilde{y}^{(\kappa)}}
\newcommand{\piik}{\pi^{(\kappa)}}
\newcommand{\tpiik}{\widetilde{\pi}^{(\kappa)}}
\allowdisplaybreaks
\begin{document}
\title{Max-min Rate Optimization of Low-Complexity  Hybrid Multi-User Beamforming Maintaining
Rate-Fairness }
\author{W. Zhu$^{1,2}$, H. D. Tuan$^2$, E. Dutkiewicz$^2$,  H. V. Poor$^3$, and L. Hanzo$^4$
\thanks{The work was supported in part by the Australian Research Council's Discovery Projects under Grant DP190102501,  in part by the U.S National Science Foundation under Grants CNS-2128448 and ECCS-2335876, in part by  the Engineering and Physical Sciences Research Council projects EP/W016605/1, EP/X01228X/1 and EP/Y026721/1 as well as of the European Research Council's Advanced Fellow Grant QuantCom (Grant No. 789028)}
	\thanks{$^1$School of Communication and Information Engineering, Shanghai University, Shanghai 200444, China
		(email: wenbozhu@shu.edu.cn); $^2$School of Electrical and Data Engineering, University of Technology Sydney, Broadway, NSW 2007, Australia (email: wenbo.zhu@student.uts.edu.au, tuan.hoang@uts.edu.au, eryk.dutkiewicz@uts.edu.au); $^3$Department of Electrical and Computer Engineering, Princeton University, Princeton, NJ 08544, USA (email: poor@princeton.edu);
		$^4$School of Electronics and Computer Science, University of Southampton, Southampton, SO17 1BJ, U.K (email: lh@ecs.soton.ac.uk) }
}
\date{}
\maketitle
\begin{abstract}
A wireless network serving multiple users in the millimeter-wave or
the sub-terahertz band by a base station is
considered. High-throughput multi-user hybrid-transmit beamforming is
conceived by maximizing the minimum rate of the users. For the sake of
energy-efficient signal transmission, the array-of-subarrays structure
is used for analog beamforming relying on low-resolution phase
shifters. We develop a convex-solver based algorithm, which
iteratively invokes a convex problem of the same beamformer size for
its solution. We then introduce the soft max-min rate objective
function and develop a scalable algorithm for its optimization. Our
simulation results demonstrate the striking fact that soft max-min
rate optimization not only approaches the minimum user rate obtained
by max-min rate optimization but it also achieves a sum rate similar
to that of sum-rate maximization. Thus, the soft max-min rate
optimization based beamforming design conceived offers a new technique
of simultaneously achieving a high individual quality-of-service for all users
and a high total network throughput.
\end{abstract}

\section{Introduction}
The millimeter-wave (mMwave) band ranging from $30$ to $300$ GHz and the
sub-Terahertz (sTHz) band ranging from  $0.1$ to  $1$ THz~\cite{Juetal21} have emerged
as the leading candidates for spectrum exploitation  in addressing the forthcoming spectrum scarcity and facilitating high-volume data delivery. These bands offer explicit advantages due to their rapidly developing advanced circuit design~\cite{RMG11,FM10,Okaetal13,AJ16,PJZ20}.

To mitigate the significant path loss experienced in the mMwave and sTHz bands, as well as to manage power consumption in circuitry, it is necessary to utilize a large number of transmit antennas (TAs) while limiting the number of radio frequency (RF) chains used for signal transmission.
Hybrid beamforming (HBF) modelled by the matrix-vector
product of analog and digital (baseband) beamforming  is considered the most promising signal processing technique for addressing these challenges.

Initially, analog beamforming (ABF) was based on a fully-connected (FC) architecture, where each RF chain was connected to all antennas. However, it necessitated an excessive number of phase shifters, even for a low number of RF chains, and thus still consumed considerable power. Recently, the
array-of-subarrays structure (AOSA)~\cite{AHRP13} has emerged as a much
more practical low-power solution for
HBF~\cite{Duetal18,Kanetal22}, where each RF
chain is connected to a subset of antennas. The AOSA also also enables the utilization of more RF chains, thereby improving the spatial diversity attained.

The HBF design has been the subject of extensive research
\cite{LL16,SY16,Chen17tvt,KHY18,Shi-18-Jun-A,TCC19,Nietal20tvt,Nasetal20TVT,Zhaetal20a,Zhaetal20tvt,Haoetal21,FMLS21,Gaoetal21,
GYSM21,QLYL22,WLLH22,Liuetal22,Cui20wcl,Lietal22twc}, with single user HBF being considered in
\cite{Chen17tvt,TCC19,Gaoetal21,QLYL22,Liuetal22} and multiuser (MU) HBF being
considered in \cite{KHY18,Haoetal21,Nietal20tvt,FMLS21,Cui20wcl,Lietal22twc,Zhaetal20a,Zhaetal20tvt}.  Due to
the computationally challenging unit modulus constraints imposed on
each entry of the ABF matrix, all these papers have only developed
heuristic procedures, which do not guarantee  convergence or
predictable performance. For instance, the authors
of~\cite{Haoetal21,FMLS21} assumed that there was no MU interference
in their ABF alternating optimization and utilized semi-definite
relaxation (SDR) in their baseband beamforming (BBF) solution. Similarly, the authors of~\cite{KHY18} utilize SDR in
both their ABF and BBF alternating optimization. It should be
mentioned that SDR is based on convex problems of excessive
dimensions. For instance, for alternating optimization of the ABF
matrix of size $64\times 4$ having $256$ decision variables as
considered in~\cite{KHY18}, the resultant SDR involves $256\times
257/2=32,896$ decision variables. Such a complex computation is clearly
beyond the capacity of existing convex-solvers. Moreover, SDR cannot
be used in alternating optimization, since it cannot generate a
feasible point.\footnote{SDR is only efficient in very limited cases,
  namely when the SDR problem has only a single solution of
  rank-one. Otherwise, it does not perform better than a very trivial
  technique~\cite{PTKN12}}

Another issue of MU beamforming is that it
is often based on sum rate (SR) maximization~\cite{SY16,Shi-18-Jun-A},
which results in zero rates for many users \cite[Table
  II]{Yuetal23tvt}. To improve the rates of all users while
maintaining computational tractability, our previous
treatise~\cite{Yuetal23tvt}  proposed maximizing the geometric mean
of the users' rates. However, the ratio of the minimum and maximum
rates \cite[Table III]{Yuetal23tvt} is still well below $0.25$, instead of approaching unity for the sake of rate-fairness.
The authors of \cite{Wuetal23} aim for maximizing the sum dirty paper coding (DPC) rate, which is capable of  providing fairer rate distributions than conventional SR maximization
\cite{Nguetal17jsac}. However, DPC is a
strictly information-theoretic concept, which cannot be implemented in
practice.

Against the above background, this is the first piece of work that
considers the HBF design problem of providing uniformly high
throughputs for all users. In contrast to other studies, we also
restrict the phase shifters to have low resolution for practical
implementation.  In a nutshell, our contributions are three-fold:
\begin{itemize}
	\item We develop a convex-solver based algorithm for HBF
          design by maximizing the users' minimum rate (MR), which
          iteratively invokes a convex problem of the same beamformer
          size to generate a gradually improved feasible point;
	\item We propose a new optimization formulation, termed as
          soft max-min rate optimization for addressing the
          computational issues encountered in high-dimensional
          nonconvex problems. Accordingly, a scalable algorithm is
          developed for their solution, which is based on a
          closed-form expression for gradually generating an improved
          feasible point;
	\item The extensive simulations show the striking benefits of
          soft max-min optimization based beamforming: its minimum
          rate (MR) is almost as high as that of max-min rate
          optimization based beamforming, and its SR performance approaches that of SR maximization based beamforming. {\color{black}Hence, utilizing soft max-min optimization yields valuable insights into identifying beneficial near-optimal solutions for the concurrent SR and MR objectives}.
\end{itemize}
In Table \ref{compare}, we boldly contrast our contributions to the related literature.

\begin{table*}[!htb]
	\centering
	\caption{Our novel contributions}
	\begin{tabular}{|l|c|c|c|c|c|c|c|}
		\hline
		\backslashbox{Contents}{Literature} & \textbf{This work} & \cite{SY16,Shi-18-Jun-A}&\cite{KHY18}&\cite{Haoetal21,FMLS21}&
		\cite{Yuetal23tvt}& \cite{Wuetal23}\\
		\hline
		Energy-efficiency & $\surd$  &  &  &  & &    \\
		\hline
		Zero rates &   &$\surd$ &  &  &  &   \\
		\hline
		Low complexity & $\surd$  & & & &  &   \\
		\hline
		Scalable computations & $\surd$&  &  & & $\surd$ &  \\
		\hline
		Algorithmic convergence&$\surd$ &  &  & & $\surd$ &  \\
		\hline
		Low-resolution ABF&$\surd$ &  &  & & $\surd$ &  \\
		\hline
		Uniformly high rates&$\surd$ &  &  & &  &  \\
		\hline
		High sum rate&$\surd$ &  &  & &  &  \\
		\hline
	\end{tabular}
	\label{compare}
\end{table*}
The remainder of the paper is organized as follows. Section II is devoted to the development of a convex-solver based algorithm for the max-min rate optimization of HBFs. Section III is dedicated to the conception of a scalable algorithm for the soft max-min rate optimization of HBFs. Section IV considers similar designs for the case of ABFs under the FC structure. Section V provides our simulations, while Section VI concludes the paper. The Appendix provides mathematical tools for the algorithmic derivations.

\emph{Notation.\;} Only the optimization variables are boldfaced; $C(0,a)$ for $a>0$ represents the set of Gaussian distributions with zero mean and power $a$;
$\angle x$ is the argument of a complex number $x$; The inner product between the matrices $X$ and $Y$ is defined by $\la X,Y\ra={\sf trace}(X^HY)$; We also use $\la X\ra$ for the trace of $X$ when $X$ is a long matrix expression. Furthermore, $[X]^2$ refers to $XX^H$, so we have $[X^H]^2=X^HX$, and $||AX||^2=\la [A^H]^2,[X]^2$ for the matrices $X$ and $A$.
For a real vector $\theta=(\theta_1,\dots, \theta_n)^T$, we define $e^{\jmath \theta}$ as the complex vector $(e^{\jmath \theta_1},\dots, e^{\jmath\theta_n})^T$, ${\sf diag}[X_i]_{i\in \clI}$ forms a matrix arranging $X_i$, $i\in \clI$ in diagonal format. For instance, ${\sf diag}[A_i]_{i=1,2}\triangleq \begin{bmatrix}A_1& 0\cr
	0 &A_2\end{bmatrix}$.

{\color{black}
\emph{Ingredient.\;} According to \cite[p. 366]{Tuybook},
a function $\bar{f}$ is said to be a tight minorant (majorant, resp.) of a function
$f$ over the domain $\mbox{dom}(f)$ at
a point $\bar{x}\in\mbox{dom}(f)$ if $f(x)\geq \bar{f}(x)\ \forall\ x\in\mbox{dom}(f)$
($f(x)\leq \bar{f}(x)\ \forall\ x\in\mbox{dom}(f)$, resp.) and
$f(\zk)=\bar{f}(\zk)$. When $\bar{f}$ is a tight minorant, $f(x_{opt})\geq f(\bar{x})$ holds for
$x_{opt}=\mbox{arg}\max_{x\in\mbox{dom}(f)} \bar{f}(x)$. When $\bar{f}$ is a tight majorant, we have $f(x_{opt})\leq f(\bar{x})$ for
$x_{opt}=\mbox{arg}\min_{x\in\mbox{dom}(f)} \bar{f}(x)$.
\section{Max-min rate optimization based HBF design}
We consider the  downlink (DL) of a base station (BS) serving {\color{black} $K$} users  indexed by
$k\in \clK\triangleq \{1, 2,\dots, K\}$. The BS is equipped with a massive $N$-antenna array,
while each user equipment (UE) $k$ has a single antenna.

For $\clN\triangleq \{1,\dots, N\}$ and $\clN_{RF}\triangleq \{1,\dots, N_{RF}\}$, where $N_{RF}$ is the number of RF chains that the BS uses for HBF, let  us assume that each RF chain is connected to only $L=N/N_{RF}$ antennas, so the phase shift based AB matrix $V_{RF}(\btheta)$  has the following AOSA structure \cite{AHRP13}:
\begin{equation}\label{sa1}
	V_{RF}(\btheta)\triangleq {\sf diag}[v_{RF}^{j}(\btheta_j)]_{j=1,\dots, N_{RF}},
\end{equation}	
with $v_{RF}^j(\btheta_j)=e^{\jmath\btheta_j}\in\mathbb{C}^L$,
for $\btheta_j\triangleq \left(\btheta_{1,j}, \dots, \btheta_{L,j}\right)^T\in\mathbb{R}^L$, which
satisfy the following discrete constraints of $b$-bit resolution
for their practical implementation \cite{KYW13}:
\begin{equation}\label{br1}
	\btheta_{\ell,j}\in \clB\triangleq \{ \nu\frac{2\pi}{2^b}, \nu=0, 1, \dots, 2^b-1\}, (\ell,j)\in\clL\times \clN_{RF},
\end{equation}
with $\clL\triangleq \{1,\dots, L\}$.
This AOSA  only needs $N$ phase shifters, so the circuit power consumption (in mW unit) is
\begin{equation}\label{poc3}
	N_{RF}\times 118 +N\times 20=\frac{N}{L}\times 118 +N\times 20,
\end{equation}	
where $118$ mW is the power consumption per RF chain \cite{DR15}, and $20$ mW is the power consumption per phase shifter \cite{Winetal13}.

Let $h_k\triangleq \begin{bmatrix}h_{k,1}&\dots&h_{k,N_{RF}}\end{bmatrix}\in\mathbb{C}^{1\times N}$ along with $h_{k,j}\in\mathbb{C}^{1\times L}$ represent the channel  between  the  BS and UE {\color{black}$k\in\clK$}, which is assumed to be known.\footnote{The reader is referred e.g. to \cite{Nietal20} and to the references therein for its efficient estimation.}

For $s_k\in C(0,1)$  being the information intended for UE $k$, which is "beamformed" by
$\bv_k^B\in\mathbb{C}^{N_{RF}}$, the signal received at UE $k$ is
\begin{eqnarray}
	y_k&=&h_k V_{RF}(\btheta)\sum_{\ell=1}^{K}\bv_\ell^Bs_{\ell}+ n_k \label{mimo}\\
	&=&\hbar_k(\btheta)\sum_{\ell=1}^{K}\bv_\ell^Bs_{\ell}+ n_k, \label{mimoe}
\end{eqnarray}
where $n_k\in C(0,\sigma)$ is the  background noise, and
\begin{eqnarray}\label{clhi}
	\hbar_k(\btheta)&\triangleq& h_kV_{RF}(\btheta)\nonumber\\
	&=&\begin{bmatrix}h_{k,1}v_{RF}^1(\btheta_1)&\dots&h_{k,N_{RF}}v_{RF}^{N_{RF}}(\btheta_{N_{RF}})\end{bmatrix}\nonumber\\
	&&\in\mathbb{C}^{1\times N_{RF}}, k\in\clK.
\end{eqnarray}
We will also use the following representations:
\begin{eqnarray}
	\hbar_k(\btheta)\bv^B_\ell&=& \sum_{j=1}^{N_{RF}} h_{k,j}v_{RF}^j(\btheta_j) \bv^B_\ell(j) \nonumber\\
	&=& \tih_{k}(\bv^B_\ell)v_{RF}(\btheta) \label{sa7}
\end{eqnarray}	
for
\begin{equation}\label{sa8}
	\begin{array}{c}
	\btheta\triangleq \begin{bmatrix}
		\btheta_1\cr
		\dots\cr
		\btheta_{N_{RF}}
	\end{bmatrix}\in\mathbb{R}^N,
	v_{RF}(\btheta)\triangleq \begin{bmatrix}
		v_{RF}^1(\btheta_1)\cr
		\dots\cr
		v_{RF}^{N_{RF}}(\btheta_{N_{RF}})
	\end{bmatrix}\in\mathbb{C}^N,  \\
\tih_{k}(\bv^B_\ell)\triangleq \begin{bmatrix}
		\bv^B_{\ell}(1)h_{k,1}&\dots&\bv^B_{\ell}(N_{RF})h_{k,N_{RF}}
	\end{bmatrix}.
\end{array}
\end{equation}	
Let the BBF matrix be defined by
\begin{equation}\label{miso2}
	\bv^B=\begin{bmatrix}\bv_1^B&\dots&\bv_{K}^B\end{bmatrix}\in\mathbb{C}^{N_{RF}\times K}.
\end{equation}
From  (\ref{mimoe}), the achievable rate of UE $k$ is defined by
\begin{equation}\label{ad4}
	r_k(\btheta,\bv^B)\triangleq \ln\left(1+\frac{|\hbar_k(\btheta)\bv^B_k|^2}{\psi_k(\btheta,\bv^B)}\right),
\end{equation}
with
\begin{equation}\label{Psii}
	\psi_k(\btheta,\bv^B)\triangleq \sum_{\ell\neq k}^{K}|\hbar_k(\btheta)\bv^B_{\ell}|^2+\sigma.
\end{equation}
Given the power budget $P$, the BS's transmit power is  constrained as
\begin{eqnarray}
	\sum_{k=1}^{K}||V_{RF}(\btheta)\bv^B_k||^2&=& \sum_{k=1}^{K} \sum_{j=1}^{N_{RF}}||v^j_{RF}(\btheta_j)\bv^B_k(j)||^2\nonumber\\
	&=&  L\sum_{k=1}^K||\bv^B_k||^2\leq P\nonumber\\
	\Leftrightarrow \sum_{k=1}^K||\bv^B_k||^2\leq P/L, \label{sa6}
\end{eqnarray}
which is  independent of $\btheta$.
We consider the following problem of max-min rate optimization:
\begin{equation}\label{basic}
	\max_{\btheta,\bv^B} f(\btheta,\bv^B)\triangleq \min_{k\in\clK}r_k(\btheta,\bv^B) \quad\mbox{s.t.}\quad  (\ref{br1}), (\ref{sa6}),
\end{equation}
which is computationally challenging due to the following complications: $(a)$ the rate function $r_k(\btheta,\bv^B)$ is nonconcave, making the optimization objective function (OF) in (\ref{basic}) both nonsmooth (nondifferentiable) and also nonconcave; $(b)$ The constraint (\ref{br1}) is
discrete, having as many as $2^{bN}$ discrete feasible points for optimization in $\btheta$.

For circumventing the issue $(b)$,  we introduce a new continuous variable
\begin{equation}\label{sa5}
	\bphi\triangleq \left(\bphi_1^T,\dots, \bphi_{N_{RF}}^T\right)^T\in\mathbb{C}^N,
	\bphi_j\triangleq\left(
	\bphi_{1,j}, \dots, \bphi_{L,j}\right)^T\in\mathbb{C}^L.
\end{equation}	
We then define
\begin{equation}\label{sa9}
	r_k(\bphi,\bv^B)\triangleq \ln\left(1+\frac{|\hbar_k(\bphi)\bv^B_k|^2}{\psi_k(\bphi,\bv^B)}\right),
\end{equation}
for
\begin{eqnarray}\label{sa5a}
	\hbar_k(\bphi)&=&h_k {\sf diag}[\bphi_j]_{j=1,\dots, N_{RF}}\nonumber\\
	&=&\begin{bmatrix}h_{k,1}\bphi_1&\dots&h_{k,N_{RF}}\bphi_{N_{RF}}\end{bmatrix},
\end{eqnarray}	
and
\begin{equation}\label{sa10}
	\psi_k(\bphi,\bv^B)\triangleq \sum_{\ell\neq k}^{K}|\hbar_k(\bphi)\bv^B_{\ell}|^2+\sigma.
\end{equation}
By defining $f(\bphi,\bv^B)\triangleq \min_{k\in\clK}r_k(\bphi,\bv^B)$,
we address the following penalized optimization for solving the problem in (\ref{basic}):
\begin{eqnarray}\label{sa11}
	\max_{\btheta,\bv^B,\bphi} f_{\gamma}(\btheta,\bphi,\bv^B)\triangleq f(\bphi,\bv^b)-\gamma||\bphi-v_{RF}(\btheta)||^2\nonumber\\ \mbox{s.t.}\quad (\ref{br1}), (\ref{sa6}),
\end{eqnarray}
where $\gamma>0$ is a penalty parameter. Note that no constraint is imposed on $\bphi$, while
the discrete variable $\btheta$ is decoupled from the max-min rate OF. {\color{black}The motivated reader
is referred to \cite[Chapter 16]{Betal06} for discussions on the efficacy of the penalty optimization methodology.}

We now propose an alternating optimization-based procedure for the solution of (\ref{sa11}). Initialized by the triplet
$(v^{B,(0)}, \phi^{(0)}, \theta^{(0)})$ feasible for (\ref{sa11}), let $(\vBk,\phik,\thetak)$ be a feasible point for (\ref{sa11}) that is found from the $(\kappa-1)$-st iteration.
\subsection{Alternating optimization in BBF}
We seek BBF $\vBko$ ensuring that $f_{\gamma}(\thetak,\phik,\vBko)> f_{\gamma}(\thetak,\phik,\vBk)$, which is the same as
\begin{eqnarray}\label{ad1}
	f(\phik,\vBko)> f(\phik,\vBk),
\end{eqnarray}
by considering the following  problem:
\begin{equation}\label{ad2}
	\max_{\bv^B} f(\phik,\bv^B)\triangleq \min_{k\in\clK}r_k(\phik,\bv^B)\quad\mbox{s.t.}\quad (\ref{sa6}).
\end{equation}
Recalling from (\ref{ad4}) and (\ref{Psii}) that $r_k(\phik,\bv^B)=\ln\left[1+|\hbar_k(\thetak)\bv^B_k|^2/\psi_k(\thetak,\bv^B)\right]$ with
\begin{equation}\label{psikb}
	\psi_k(\thetak,\bv^B)=\sum_{\ell\neq k}^{K}|\hbar_k(\thetak)\bv^B_{\ell}|^2+\sigma,
\end{equation}
as well as by applying the inequality (\ref{inv2}) for $\bar{x}=\xk_{k}\triangleq \hbar_k(\phik)\vBk_k$ and
$\bar{y}=\yk_{k}\triangleq \psi_k(\phik,\vBk)$, we obtain the following tight concave quadratic
minorant of $r_k(\phik,\bv^B)$ at $\vBk$:
\begin{equation}\label{ad3}
	\rk_k(\bv^B)\triangleq\alphak_k+2\Re\{\ak_k\bv_k^B \}-\betak_k\la [\hbar^H_k(\phik)]^2,\sum_{\ell=1}^{K}[\bv_\ell^B]^2\ra,
\end{equation}
with
$\alphak_k\triangleq r_k(\phik,\vBk)-|\xk_{k}|^2/\yk_{k}
-\sigma \betak_k$,
$\ak_k\triangleq  (\xk_{k})^*\hbar_k(\phik)/\yk_{k}$,
$\betak_k\triangleq 1/\yk_{k}-1/(\yk_{k}+|\xk_{k}|^2)$.

We thus solve the following convex problem of minorant maximization of (\ref{ad2}) to generate $\vBko$ ensuring (\ref{ad1}):
\begin{equation}\label{ad4m}
	\max_{\bv^B}\fk_B(\bv^B)\triangleq \min_{k\in\clK} \rk_k(\bv^B)\quad\mbox{s.t.}\quad  (\ref{sa6}).
\end{equation}
\subsection{Alternating optimization in $\bphi$}
We seek  $\phiko$ for ensuring that
\begin{eqnarray}\label{phi1}
	f_{\gamma}(\thetak,\phiko,\vBko)> f_{\gamma}(\thetak,\phik,\vBko)
\end{eqnarray}
by considering the following problem:
\begin{eqnarray}\label{saphi2}
	&&\max_{\bphi}f_{\gamma}(\thetak,\bphi,\vBko)\nonumber\\
	&\triangleq& \min_{k\in\clK}r_k(\bphi,\vBko)-\gamma ||\bphi-v_{RF}(\thetak)||^2.
\end{eqnarray}
By recalling from (\ref{sa9}) and (\ref{sa7})
that $r_k(\bphi,\vBko)\triangleq \ln\left[1+| \tih_{k}(\vBko_k)\bphi|^2/\psi_k(\bphi,\vBko) \right]$ with
\begin{equation}\label{psikz}
	\psi_k(\bphi,\vBko)=\sum_{\ell\neq k}| \tih_{k}(\vBko_\ell)\bphi|^2+\sigma,
\end{equation}
and by applying the inequality (\ref{inv2}) of the Appendix for $\bar{x}=\txk_{k}\triangleq \tih_{k}(\vBko_k)\phik$ and
$\bar{y}=\tyk_{k}\triangleq \psi_k(\phik,\vBko)$, we obtain the following tight minorant of $r_k(\bphi,\vBko)$ at $\phik$:
\begin{equation}\label{sphi3}
	\trk_k(\bphi)\triangleq\talphak_k+2\Re\{\tak_k\bphi \}
	-\tbetak_k \la \sum_{\ell=1}^{K}[\tih^H_{k}(\vBko_\ell)]^2,[\bphi]^2\ra,
\end{equation}
with	$\talphak_k\triangleq r_k(\phik,\vBko)-| \txk_{k}|^2/\tyk_{k}
-\sigma\tbetak_{k}$,
$\tak_k\triangleq \ds\frac{(\txk_{k})^*}{\tyk_{k}}\tih_{k}(\vBko_k)$,
$\tbetak_k\triangleq 1/\tyk_{k}-1/\left(\tyk_{k}+|\txk_{k}|^2 \right)$.

We thus solve the following convex problem of  minorant maximization formulated in (\ref{saphi2})
for generating $\phiko$ satisfying (\ref{phi1}):
\begin{equation}\label{sphi4}
	\max_{\bphi}\fk_{\gamma,z}\triangleq \min_{k\in\clK}\trk_k(\bphi)-\gamma||\bphi-v_{RF}(\thetak)||^2.
\end{equation}
\subsection{Alternating optimization in ABF}
To seek  $\thetako$ for ensuring that $f_{\gamma}(\thetako,\phiko,\vBko)> f_{\gamma}(\thetak,\phiko,\vBko)$, which is the same as
\begin{equation}\label{the1}
	||\phiko-v_{RF}(\thetako)||^2<||\phiko-v_{RF}(\thetak)||^2,
\end{equation}
we consider the problem $\min_{\btheta} ||\phiko-v_{RF}(\btheta)||^2\quad\mbox{s.t.}\quad (\ref{br1})$,
which admits the following closed-form solution:
\begin{equation}\label{the3}
	\thetako_{\ell,j}=\lfloor\angle\phiko_{\ell,j}\rceil_{b}, (\ell,j)\in\clL\times\clN_{RF}.
\end{equation}
Here and after,  $\lfloor\alpha\rceil_b$ is the $b$-bit rounded version of  $\alpha\in [0,2\pi)$ defined by $\lfloor\alpha\rceil_b=\nu_{\alpha}\frac{2\pi}{2^b}$
with $\nu_{\alpha}\triangleq \mbox{arg}\min_{\nu'\in\{\nu,\nu+1\}}\left|\nu' \frac{2\pi}{2^b}-\alpha\right|$, where $\nu$ is selected for satisfying that
$\alpha\in [\nu \frac{2\pi}{2^b}, (\nu+1)\frac{2\pi}{2^b}]$. If $\nu_{\alpha}= 2^b$
we reset it to $\nu_{\alpha}= 0$.
\subsection{Max-min rate optimization and its convergence}
Algorithm \ref{alg1} summarizes the computational procedure iterating by solving the convex problems (\ref{ad4m}) as well as (\ref{sphi4}), and the closed-form (\ref{the3}) to generate a sequence $\{(\phik, \thetak, \vBk)\}$
of improved feasible points for (\ref{sa11}), because we have
$f_{\gamma}(\phiko,\thetako,\vBko)> f_{\gamma}(\phik,\thetak,\vBk)$ by
(\ref{ad1}), (\ref{phi1}), and (\ref{the1}). This sequence is convergent according to Cauchy's theorem. Moreover, for a sufficient large $\gamma$, we have $||\phik-v_{RF}(\thetak)||\rightarrow 0$,
so $(\thetak,\vBk)$ represents an optimized solution of the max-min rate optimization problem (\ref{basic}).

\begin{algorithm}[!t]
	\caption{Max-min rate optimization-based HBF algorithm } \label{alg1}
	\begin{algorithmic}[1]
		\State \textbf{Initialization:} Initialize a feasible
point $(\theta^{(0)}, \phi^{(0)}, v^{B,(0)})$ for (\ref{sa11}).
		\State \textbf{Repeat until convergence of the objective function in  (\ref{sa11}):} Generate  $\vBko$ by
		solving the convex problem (\ref{ad4m}),
		$\phiko$  by solving the convex problem (\ref{sphi4}), and $\thetako$ by (\ref{the3}).
		Reset $\kappa:=\kappa+1$.
		\State \textbf{Output} $(\theta^{opt}, \tilde{v}^{B,opt})=(\thetak,\vBk)$.
	\end{algorithmic}
\end{algorithm}
\subsection{SR maximization based HBF algorithm}
Instead of the problem (\ref{basic}) of max-min optimization, we now consider the following problem of SR maximization:
\begin{equation}\label{subasic}
	\max_{\btheta,\bv^B} g(\btheta,\bv^B)\triangleq \sum_{k=1}^Kr_k(\btheta,\bv^B) \quad\mbox{s.t.}\quad  (\ref{br1}), (\ref{sa6}),
\end{equation}
which is addressed based on the following problem of penalized optimization:
\begin{eqnarray}\label{susa11}
	\max_{\btheta,\bv^B,\bphi} g_{\gamma}(\btheta,\bphi,\bv^B)\triangleq \sum_{k=1}^K r_k(\bphi,\bv^B)-\gamma||\bphi-v_{RF}(\btheta)||^2\nonumber\\ \mbox{s.t.}\quad (\ref{br1}), (\ref{sa6}).
\end{eqnarray}
Initialized by $(z^{(0)}, \phi^{(0)}, \theta^{(0)})$ feasible for (\ref{susa11}), let $(\zk,\phik,\thetak)$ be a feasible point for (\ref{susa11}) that is found from the $(\kappa-1)$-st iteration. The alternating optimization at the $\kappa$-th iteration proceeds as follows.
\subsubsection{Alternating optimization in BBF}
Similarly to (\ref{ad4m}), we generate $\vBko$ by solving the problem
\begin{equation}\label{suad4m}
	\max_{\bv^B} \sum_{k=1}^K \rk_k(\bv^B)\quad\mbox{s.t.}\quad  (\ref{sa6}),
\end{equation}
where $\rk_k$ is defined from (\ref{ad3}).  By taking into account that
$\sum_{k=1}^K \rk_k(\bv^B)=\alphak+\sum_{k=1}^K2\Re\{\ak_k\bv_k^B \}-\sum_{k=1}^K\la\Xik,[\bv_k^B]^2\ra	$
with $\alphak\triangleq \sum_{k=1}^K\alphak_k$ and	$\Xik=\sum_{k=1}^L\betak_k[\hbar^H_k(\phik)]^2$, the problem (\ref{suad4m}) admits the closed-form solution of
\begin{equation}\label{suad6}
	\vBko_k=\begin{cases}\begin{array}{ll}(\Xik)^{-1}(\ak_k)^H\\ \mbox{if}\quad \sum_{k=1}^K||(\Xik)^{-1}(\ak_k)^H||^2\leq P/L,\cr
			(\Xik+\lambda I_{N_{RF}})^{-1}(\ak_k)^H\\ \mbox{otherwise},
		\end{array}
	\end{cases}
\end{equation}	
where $\lambda>0$ is found by bisection, so that $\sum_{k=1}^K||	(\Xik+\lambda I_{N_{RF}})^{-1}(\ak_k)^H||^2=P/L$.
\subsubsection{Alternating optimization in $\bphi$}
Like in (\ref{sphi4}), we  generate $\phiko$ by solving the problem
\begin{equation}\label{susphi4}
	\max_{\bphi} \sum_{k=1}^K\trk_k(\bphi)-\gamma||\bphi-v_{RF}(\thetak)||^2,
\end{equation}
where $\trk_k(\bphi)$ is defined in (\ref{sphi3}). By presenting $\sum_{k=1}^K\trk_k(\bphi)-\gamma||\bphi-v_{RF}(\thetak)||^2=\balphak+2\Re\{\tak\bphi \}-\la\tXik,[\bphi]^2\ra$ with $\balphak\triangleq \sum_{k=1}^K\talphak_k-\gamma N$, and
\begin{equation}\label{susphi5}
	\begin{array}{c}
	\tak\triangleq \ds\sum_{k=1}^K\tak_k+\gamma [v_{RF}(\thetak)]^H, \\
	\tXik\triangleq \ds\sum_{k=1}^K\tbetak_k \sum_{\ell=1}^{K}[\tih^H_{k}(\vBko_\ell)]^2+\gamma I_N,
	\end{array}
\end{equation}	
the problem (\ref{susphi4}) admits the closed-form solution of
\begin{equation}\label{susphi6}
	\phiko=(\tXik)^{-1}(\tak)^H.
\end{equation}	
\subsubsection{Algorithm }
It may now be seen that alternating optimization in $\btheta$ is based on the closed-form (\ref{the3}). As such, Algorithm \ref{salg1} constructed for solving problem (\ref{susa11}) is of scalable
complexity, {\color{black} with the total computational complexity of each iteration being on the order of ${\cal O}(N_{RF}K)+{\cal O}(N)$}.

\begin{algorithm}[!t]
	\caption{Scalable SR maximization-based HBF algorithm } \label{salg1}
	\begin{algorithmic}[1]
		\State \textbf{Initialization:} Initialize $(\theta^{(0)}, \phi^{(0)}, v^{B,(0)})$.
		\State \textbf{Repeat until convergence of the objective function in  (\ref{susa11}):} Generate  $\vBko$ by
		(\ref{suad6}),
		and $\phiko$  by  (\ref{susphi6}), and $\thetako$ by (\ref{the3}).
		Reset $\kappa:=\kappa+1$.
		\State \textbf{Output} $(\theta^{opt}, \tilde{v}^{B,opt})=(\thetak,\vBk)$.
	\end{algorithmic}
\end{algorithm}
\section{Soft max min rate optimization based HBF design}
The total computational complexity of the convex problems (\ref{ad4m}) and (\ref{sphi4}) that are solved
at each iteration of Algorithm \ref{alg1} is  on the order of  ${\cal O}[(N_{RF}K)^3]+{\cal O}(N^3)$, which is high, because $N$ is large. This motivates  us in this section to develop another technique of finding the best MR by scalable computation.

One has
\begin{eqnarray}
	&&\max_{\btheta,\bv^B}\min_{k\in\clK}r_k(\btheta,\bv^B)\nonumber\\
	&\Leftrightarrow&
	\max_{\btheta,\bv^B}\min_{k\in\clK}\ln\left(1+ \frac{1}{c}\frac{|\hbar_k(\btheta)\bv^B_k|^2}{\psi_k(\btheta,\bv^B)} \right)\label{dev0}\\
	&\Leftrightarrow&\max_{\btheta,\bv^B}\left[-\max_{k\in\clK}\ln\left(1+ \frac{|\hbar_k(\btheta)\bv^B_k|^2}{c\psi_k(\btheta,\bv^B)}  \right)^{-1}\right],\label{dev01}
\end{eqnarray}
while
\begin{eqnarray}
	&&\max_{k\in\clK}\ln\left(1+ \frac{|\hbar_k(\btheta)\bv^B_k|^2}{c\psi_k(\btheta,\bv^B)}  \right)^{-1}\nonumber\\ &\leq&\ln\left(\sum_{k=1}^{K}\left(1+ \frac{|\hbar_k(\btheta)\bv^B_k|^2}{c\psi_k(\btheta,\bv^B)}\right)^{-1}\right)\label{dev1}\\
	&=&\ln\left(\frac{\sum_{k=1}^{K}\left(1+ \frac{|\hbar_k(\btheta)\bv^B_k|^2}{c\psi_k(\btheta,\bv^B)}\right)^{-1}}{K}\right)
	+\ln K\label{dev2}\\
	&\leq&\max_{k\in\clK}\ln\left(1+ \frac{|\hbar_k(\btheta)\bv^B_k|^2}{c\psi_k(\btheta,\bv^B)}  \right)^{-1}
	+\ln K.\label{dev4}
\end{eqnarray}
Note that for sufficiently small $c$, $\ln K$ is very small compared to the absolute value of the
LHS of (\ref{dev1}). In other words, by choosing small enough $c$, the LHS of (\ref{dev1}) can be approximated with arbitrary tolerance
by the right-hand side (RHS) of (\ref{dev2}), which is $\ln\pi_c(\btheta,\bv^B)$ for
\begin{eqnarray}\label{nr1}
	\pi_c(\btheta,\bv^B)\triangleq \sum_{k=1}^{K}\left(1- \frac{|\hbar_k(\btheta)\bv^B_k|^2}{|\hbar_k(\btheta)\bv^B_k|^2+c\psi_k(\btheta,\bv^B)}\right).
\end{eqnarray}
Instead of the max-min optimization problem (\ref{basic}), we thus consider the following problem referred to as the soft max-min optimization problem:
\begin{equation}\label{sbasic0}
	\max_{\btheta,\bv^B}\ [-\ln\pi_c(\btheta,\bv^B)] \quad\mbox{s.t.}\quad  (\ref{br1}), (\ref{sa6}),
\end{equation}
which is equivalent to the problem
\begin{equation}\label{sbasic}
	\min_{\btheta,\bv^B}\ \ln\pi_c(\btheta,\bv^B) \quad\mbox{s.t.}\quad  (\ref{br1}), (\ref{sa6}).
\end{equation}
We then address its solution  by  the following problem of penalized optimization:
\begin{eqnarray}\label{ssa1}
	\min_{\btheta,\bv^B,\bphi} f_{\gamma,c}(\btheta,\bphi,\bv^B)\triangleq \ln\pi_c(\btheta,\bv^B) +\gamma||\bphi-v_{RF}(\btheta)||^2\nonumber\\ \mbox{s.t.}\quad (\ref{br1}), (\ref{sa6}),
\end{eqnarray}
where $\gamma>0$ is a penalty parameter, and
\begin{equation}\label{sbasicf}
	\pi_c(\bphi,\bv^B)\triangleq \sum_{k=1}^{K}\left(1- \frac{|\hbar_k(\bphi)\bv^B_k|^2}{|\hbar_k(\bphi)\bv^B_k|^2+c\psi_k(\bphi,\bv^B)}\right),
\end{equation}
with $\hbar_k(\bphi)$ and $\psi_k(\bphi,\bv^B)$ defined from (\ref{sa5a}) and (\ref{sa10}).

We now propose an alternating optimization-based procedure for the solution of (\ref{ssa1}). Initialized by
$(v^{B, (0)}, \phi^{(0)}, \theta^{(0)})$ feasible for (\ref{sa11}), let $(\vBk,\phik,\thetak)$ be a feasible point for (\ref{ssa1}) that is found from the $(\kappa-1)$-st iteration.
\subsection{Alternating optimization in BBF}
We seek BBF $\vBko$ ensuring that $f_{\gamma,c}(\thetak,\phik,\vBko)< f_{\gamma,c}(\thetak,\phik,\vBk)$, which is the same as
\begin{eqnarray}
	\ln\pi_c(\phik,\vBko)<\ln\pi_c(\phik,\vBk), \label{sad1}
\end{eqnarray}
by considering the following  problem:
\begin{equation}\label{ssa2}
	\min_{\bv^B} \ln\piik_c(\bv^B)\quad
	\mbox{s.t.}\quad  (\ref{sa6}),
\end{equation}
where we have
\begin{eqnarray}
\piik_c(\bv^B)&\triangleq& \pi_c(\phik,\bv^B)\nonumber\\
&=&\sum_{k=1}^{K}\left(1- \frac{|\hbar_k(\phik)\bv^B_k|^2}{|\hbar_k(\phik)\bv^B_k|^2+c\psi_k(\phik,\bv^B)}\right)\nonumber
\end{eqnarray}
with $\psi_k(\phik,\bv^B)$ defined from (\ref{psikb}). Applying the inequality (\ref{ap6}) of the Appendix for
$\bar{x}_k=\xk_{k}\triangleq \hbar_k(\phik)\vBk_k$ and $\bar{y}_k=\yk_{k}\triangleq \psi_k(\phik,\vBk)$
yields the following tight majorant of $\ln\piik_c(\bv^B)$ at $\vBk$:
\begin{align}
	\rhok(\bv^B)\triangleq&
	\ak-2\sum_{k=1}^K\dk_k\Re\{(\xk_{k})^*\hbar_k(\phik)\bv^B_k\}\nonumber\\
	&+\sum_{k=1}^K\ck_k\left(c\sum_{\ell\neq k}|\hbar_k(\phik)\bv^B_{\ell}|^2 \right.\nonumber\\
	&\left.+|\hbar_k(\phik)\bv^B_k|^2\right)\nonumber\\
	=&\ak_b-2\sum_{k=1}^K\Re\{\bk_k\bv^B_k\}+\sum_{k=1}^K\la\Ck_k,[\bv^B_k]^2\ra,\label{sad4}
\end{align}
where
\begin{eqnarray}
	\ak\triangleq \fk_{sf,b}(\vBk)
	+\sum_{k=1}^K\dk_k|\xk_{k}|^2+c\sigma\sum_{k=1}^K\ck_k,\label{sad5}\\
	\dk_k\triangleq\frac{\left(c\yk_{k}+|\xk_{k}|^2\right)^{-1}}{\piik_c(\vBk)},
	\ck_k\triangleq\dk_k\frac{|\xk_{k}|^2}{c\yk_{k}+|\xk_{k}|^2},\label{sad7}
\end{eqnarray}
and
\begin{equation}\label{sad8}
	\begin{array}{c}
		\bk_k\triangleq \dk_k (\hbar_k(\phik)\vBk_k)^*\hbar_k(\phik),\\
		\Ck_k\triangleq\ds c\sum_{\ell\in\clK\setminus\{k\}}\ck_{\ell}[\hbar^H_{\ell}(\phik)]^2+
		\ck_k[\hbar^H_{k}(\phik)]^2.
	\end{array}
\end{equation}
We thus solve the following problem of majorant minimization to generate $\vBko$ ensuring (\ref{sad1}):
\begin{equation}\label{ssa3}
	\min_{\bv^B}\ \rhok(\bv^B)\quad\mbox{s.t.}\quad (\ref{sa6}),
\end{equation}
which admits the closed-form solution of
\begin{equation}\label{ssa4}
	\vBko_k=\begin{cases}\begin{array}{ll}(\Ck_k)^{-1}(\bk_k)^H\\
			\mbox{if}\quad \ds\sum_{k=1}^K ||(\Ck_k)^{-1}(\bk_k)^H||^2\leq P/L,\cr
			(\Ck_k+\lambda I_{N_{RF}} )^{-1}(\bk_k)^H\\ \mbox{otherwise},
		\end{array}
	\end{cases}
\end{equation}
where $\lambda>0$ is found by bisection, so that	$\sum_{k=1}^K||(\Ck_k+\lambda I_{N_{RF}} )^{-1}(\bk_k)^H||^2=P/L$.
\subsection{Alternating optimization in $\bphi$}
We seek  $\phiko$ for ensuring that
\begin{eqnarray}
	f_{\gamma,c}(\thetak,\phiko,\vBko)< f_{\gamma,c}(\thetak,\phik,\vBko),\label{sphi1}
\end{eqnarray}
by considering the following problem:
\begin{equation}\label{ssa10}
	\min_{\bphi}\ [\ln\tpiik_c(\bphi)+\gamma ||\bphi-v_{RF}(\thetak)||^2],
\end{equation}
where
\begin{eqnarray}
\tpiik_c(\bphi)&\triangleq& \pi_c(\bphi,\vBko)\nonumber\\
&=&\sum_{k=1}^{K}\left(1- \frac{ |\tih_k(\vBko_k)\bphi|^2}{|\tih_k(\vBko_k)\bphi+c\psi_k(\bphi,\vBko)}\right)\nonumber
\end{eqnarray}
with
$\psi_k(\bphi,\vBko)$ defined from (\ref{psikz}).
Applying the inequality (\ref{ap6}) for
$
\bar{x}_k=\txk_k\triangleq \tih_k(\vBko_{k})\phik$, and $\bar{y}_k=\tyk_k\triangleq \psi_k(\phik,\vBko)$
yields the following tight majorant of $\ln\tpiik_c(\bphi)$ at $\phik$:
\begin{align}
	\trhok(\bphi)\triangleq&
	\tak-2\sum_{k=1}^K\tdk_k\Re\{(\txk_k)^*\tih_k(\vBko_{k})\bphi\}\nonumber\\
	&+\sum_{k=1}^K\tck_k\left(c\sum_{\ell\neq k}|\tih_k(\vBko_{\ell})\bphi|^2\right.\nonumber\\
	&\left.+|\tih_k(\vBko_{k})\bphi|^2\right)\nonumber\\
	=&\tak-2\Re\{\tbk\bphi\}+\la\tCk,[\bphi]^2\ra,\label{ssad4}
\end{align}
where we have
\begin{eqnarray}
	\tak\triangleq \fk_{sf,z}(\phik)
	-\sum_{k=1}^K\tdk_k|\txk_k|^2-c\sigma\sum_{k=1}^K\tck_k,\label{ssad5}\\
	\tdk_k\triangleq \frac{\left(c\tyk_k+|\txk_k|^2\right)^{-1}}{\tpiik_c(\bphi)},
	\tck_k\triangleq\tdk_k\frac{|\txk_k|^2}{c\tyk_k+|\txk_k|^2},\label{ssad7}
\end{eqnarray}
and
\begin{equation}\label{ssad8}
	\begin{array}{c}
		\tbk\triangleq \sum_{k=1}^K\left( \tdk_k (\tih_k(\vBko_{k})\phik)^*\tih_k(\vBko_{k})\right),\\
		\tCk\triangleq\begin{aligned}[t]&\ds \sum_{k=1}^K\tck_{k}\left( c\sum_{\ell\in\clK\setminus\{k\}}[\tih^H_k(\vBko_{\ell})]^2\right.\\
		&\left.+[\tih^H_k(\vBko_{k})]^2\right).\end{aligned}
	\end{array}
\end{equation}
We thus solve the following problem of majorant minimization of (\ref{ssa10}) to generate $\phiko$ ensuring (\ref{sphi1}):
\begin{equation}\label{ssa11}
	\min_{\bphi} \trhok(\bphi)+\gamma ||\bphi-v_{RF}(\thetak)||^2
\end{equation}
which admits the closed-form solution of
\begin{equation}\label{ssa12}
	\phiko=(\tCk+\gamma I_{N})^{-1}\left[(\bk)^H+\gamma v_{RF}(\thetak)\right].		
\end{equation}
\subsection{Alternating optimization  in ABF}
Generate  $\thetako$ according (\ref{the3}).
\subsection{Soft max-min rate optimization and its convergence}
Algorithm \ref{alg2} summarizes the computational procedure iterating by evaluating the closed-form expressions of (\ref{ssa4}), (\ref{ssa12}), and (\ref{the3}) to generate a sequence $\{(\phik, \thetak, \vBk)\}$
of improved feasible points for (\ref{ssa1}), because
$f_{\gamma,c}(\phiko,\thetako,\vBko)<f_{\gamma,c}(\phik,\thetak,\vBk)$ by
(\ref{sad1}), (\ref{sphi1}), and (\ref{the1}). This sequence is convergent by Cauchy's theorem. Moreover, for a sufficient large $\gamma$, we have $||\phik-v_{RF}(\thetak)||\rightarrow 0$,
so $(\thetak,\vBk)$ represents an optimized solution for the soft max-min rate optimization problem (\ref{sbasic0})/(\ref{sbasic}).
{\color{black} The total computational complexity of each iteration is on the order of ${\cal O}(N_{RF}K)+{\cal O}(N)$.}
\begin{algorithm}[!t]
	\caption{Scalable soft max-min rate optimization based HBF algorithm } \label{alg2}
	\begin{algorithmic}[1]
		\State \textbf{Initialization:} Initialize $(\theta^{(0)}, \Phi^{(0)}, v^{B,(0)})$.
		\State \textbf{Repeat until convergence of the objective function in  (\ref{ssa1}):} Generate  $\vBko$ by
		(\ref{ssa4}), $\phiko$  by (\ref{ssa12}), and $\thetako$ by (\ref{the3}).
		Reset $\kappa:=\kappa+1$.
		\State \textbf{Output} $(\theta^{opt}, v^{B,opt})=(\thetak,\vBk)$.
	\end{algorithmic}
\end{algorithm}
\section{Baseline performance of fully-connected RF chains}
To show the advantage of AOSA we have to compare its performance to that of HBF using
FC-based ABF. For the full connection of each RF chain,  let
$\btheta\triangleq [\btheta_{n,j}]_{(n,j)\in\clN\times \clN_{RF}}\in [0,2\pi)^{N\times N_{RF}}$
be the phase shift matrix. Instead of the diagonal structure (\ref{sa1}), the FC ABF matrix is
`structure-free', formulated as:
\begin{equation}\label{vrf}
	V_{RF}(\btheta)\triangleq [e^{\jmath\btheta_{n,j}}]_{(n,j)\in\clN\times \clN_{RF}}.
\end{equation}
For
\begin{equation}\label{base1}
	\hbar_k(\btheta)\triangleq h_kV_{RF}(\btheta)\in\mathbb{C}^{1\times N_{RF}}, k\in\clK,
\end{equation}
the achievable rate of UE $k$  is defined by (\ref{ad4})-(\ref{Psii}), while the transmit constraint is
\begin{equation}\label{pc1}
	\sum_{k=1}^{K}||V_{RF}(\btheta)\bv^B_k||^2= \sum_{k=1}^{K} \la  [V^H_{RF}(\btheta)]^2,[\bv^B_k]^2 \ra\leq P,
\end{equation}
which is dependent on $\btheta$,  unlikely (\ref{sa6}). Our AOSA-related discussions of the previous sections are still relevant for FC, albeit with some more transforms involved in deriving the analytical forms of
${\sf vect} [V_{RF}(\btheta)]$ to find closed-form based solutions.
\subsection{Max-min rate optimization based design}
Similarly to (\ref{sa11}), we address the problem of max-min rate optimization via the following
problem of penalized optimization:
\begin{subequations}\label{bmm1}
	\begin{eqnarray}
		\max_{\btheta,\bv^B,\bPhi}\ [\min_{k=1, \dots, K}r_k(\bPhi,\bv^B)-\gamma||\bPhi-V_{RF}(\btheta)||^2]\ \mbox{s.t.}\ (\ref{br1}),\label{bmm1a}\\
		\sum_{k=1}^{K} \la  [\bPhi^H]^2,[\bv^B_k]^2 \ra\leq P,\label{bmm1b}
	\end{eqnarray}
\end{subequations}
where 	$\bPhi\in \mathbb{C}^{N\times N_{RF}}$ is the new variable,
and then we define	$r_k(\bPhi,\bv^B)\triangleq \ln\left(1+\frac{|\hbar_k(\bPhi)\bv^B_k|^2}{\psi_k(\bPhi,\bv^B)}\right)$ for
$\hbar_k(\bPhi)\triangleq h_k\bPhi\in\mathbb{C}^{N_R\times N_{RF}}$, and
$\psi_k(\bPhi,\bv^B)\triangleq \sum_{\ell\neq k}^{K}|\hbar_k(\bPhi)\bv^B_{\ell}|^2+\sigma$,
and $\gamma>0$ is a penalty parameter.

We briefly present an alternating optimization-based procedure for the solution of (\ref{bmm1}). Initialized by
$(v^{B,(0)}, \Phi^{(0)}, \theta^{(0)})$ feasible for (\ref{bmm1}), let $(\vBk,\Phik,\thetak)$ be a feasible point for (\ref{bmm1}) that is found from the $(\kappa-1)$-st iteration.
\subsubsection{Alternating optimization in BBF}
$\vBko$ is generated by solving the convex problem of
\begin{equation}\label{bad4m}
	\max_{\bv^B} \min_{k=1,\dots, K} \rk_k(\bv^B)\quad\mbox{s.t.}\quad  \sum_{k=1}^{K} \la  [(\Phik)^H]^2,[\bv^B_k]^2 \ra\leq P,
\end{equation}
where  $\rk_k(\bv^B)$ is a  tight concave quadratic
minorant of $r_k(\Phik,\bv^B)$ at $\vBk$ defined by
\begin{equation}\label{bad3}
	\rk_k(\bv^B)\triangleq\alphak_k+2\Re\{\ak_k\bv_k^B \}-\betak_k\la [\hbar^H_k(\Phik)]^2,\sum_{\ell=1}^{K}[\bv_\ell^B]^2\ra,
\end{equation}
with	$\alphak_k\triangleq r_k(\Phik,\vBk)-|\xk_k|^2/\yk_k
-\sigma \betak_k$,
$\ak_k\triangleq  (\xk_k)^*\hbar_k(\Phik)/\yk_k$,
$\betak_k\triangleq 1/\yk_k-1/(\yk_k+|\xk_k|^2)$
for $\xk_k\triangleq \hbar_k(\Phik)\vBk_k$, and $\yk_k\triangleq \psi_k(\Phik,\vBk)$.
\subsubsection{Alternating optimization in $\bPhi$}
$\Phiko$ is generated by solving the following convex problem:
\begin{eqnarray}\label{bphi4}
	\max_{\bPhi} \min_{k=1, \dots, K}\trk_k(\bPhi)-\gamma||\bPhi-V_{RF}(\thetak)||^2\nonumber\\ \mbox{s.t.}\quad
	\la[\bPhi^H]^2,\sum_{k=1}^{K}[\vBko_k]^2 \ra\leq P,
\end{eqnarray}
where $\trk_k(\bPhi)$ is a tight concave minorant of $r_k(\bPhi,\vBko)$ at $\Phik$ defined by
\begin{eqnarray}\label{bphi3}
	\trk_k(\bPhi)&\triangleq&\talphak_k+2\Re\{\la \tAk_k\bPhi\ra \}\nonumber\\
	&&-\tbetak_k\la h^H_kh_k, \bPhi(\sum_{\ell=1}^{K}[\vBko_\ell]^2)\bPhi^H\ra,
\end{eqnarray}
with $\talphak_k\triangleq r_k(\Phik,\vBko)-|\txk_k|^2/\tyk_k
-\sigma\tbetak_k$,
$\tAk_k\triangleq \ds\frac{(\txk_k)^*}{\tyk_k}\vBko_kh_k$,
$\tbetak_k\triangleq 1/\tyk_k-1/\left(\tyk_k+|\txk_k|^2\right)$,
for $\txk_k\triangleq \hbar_k(\Phik)\vBko_k$ and $\tyk_k\triangleq \psi_k(\Phik,\vBko)$.
\subsubsection{Alternating optimization in ABF}
$\thetako$ is generated according to the following formula:
\begin{equation}\label{bthe3}
	\thetako_{n,j}=\lfloor\angle\Phiko(n,j)\rceil_{b}, (n,j)\in\clN\times\clN_{RF}.
\end{equation}
\subsubsection{Algorithm}
Like Algorithm \ref{alg1}, Algorithm \ref{falg1} also generates a sequence of gradually improved feasible points for
(\ref{bmm1}), so its convergence is guaranteed by Cauchy's theorem. The total computational complexity of
the convex problems (\ref{bad4m}) and (\ref{bphi4}) is ${\cal O}(N_{RF}^3K^3)+{\cal O}(N_{RF}^3N^3)$.

\begin{algorithm}[!t]
	\caption{Max-min rate optimization based FC HFB algorithm } \label{falg1}
	\begin{algorithmic}[1]
		\State \textbf{Initialization:} Initialize $(\theta^{(0)}, \Phi^{(0)}, v^{B,(0)})$.
		\State \textbf{Repeat until convergence of the objective function in  (\ref{bmm1}):} Generate  $\vBko$ by
		solving the convex problem (\ref{bad4m}),
		and $\Phiko$  by solving the convex problem (\ref{bphi4}), and $\thetako$ by (\ref{bthe3}).
		Reset $\kappa:=\kappa+1$.
		\State \textbf{Output} $(\theta^{opt}, \tilde{v}^{B,opt})=(\thetak,\vBk)$.
	\end{algorithmic}
\end{algorithm}
\subsection{SR maximization-based design}
Similarly to (\ref{susa11}), the problem of SR maximization is addressed
via the following problem of penalized optimization:
\begin{equation}\label{sumbmm1}
	\max_{\btheta,\bv^B,\bPhi}\ [\sum_{k=1}^Kr_k(\bPhi,\bv^B)-\gamma||\bPhi-V_{RF}(\btheta)||^2]\quad \mbox{s.t.}\quad (\ref{br1}), (\ref{pc1}).
\end{equation}
\subsubsection{Alternating optimization in  BBF}
$\vBko$ is generated by solving the problem of
\begin{equation}\label{sumbad4m}
	\max_{\bv^B} \sum_{k=1}^K \rk_k(\bv^B)\quad\mbox{s.t.}\quad  \sum_{k=1}^{K} \la  [(\Phik)^H]^2,[\bv^B_k]^2 \ra\leq P,
\end{equation}
with  $\rk_k(\bv^B)$ defined from (\ref{bad3}). By expressing
\[
\sum_{k=1}^K \rk_k(\bv^B)=\alphak+\sum_{k=1}^K2\Re\{\ak_k\bv^B_k\}-\sum_{k=1}^K\la \Xik,[\bv^B_k]^2\ra,
\] with $\alphak\triangleq \sum_{k=1}^K\alphak_k$ and
$	\Xik\triangleq \sum_{k=1}^k\betak_k[\hbar^H_k(\Phik)]^2$,
the problem (\ref{sumbad4m}) admits the closed-form solution of
\begin{equation}\label{sumsuad6}
	\vBko_k=\begin{cases}\begin{array}{ll}(\Xik)^{-1}(\ak_k)^H\\ \mbox{if}\quad \sum_{k=1}^K||\Phik(\Xik)^{-1}(\ak_k)^H||^2\leq P,\cr
			(\Xik+\lambda [(\Phik)^H]^2)^{-1}(\ak_k)^H\\ \mbox{otherwise},
		\end{array}
	\end{cases}
\end{equation}	
where $\lambda>0$ is found by bisection, so that $||\Phik	(\Xik+\lambda [(\Phik)^H]^2)^{-1}(\ak_k)^H||^2=P$.
\subsubsection{Alternating optimization in $\bPhi$}
$\Phiko$ is generated by solving the following problem:
\begin{eqnarray}\label{sumbbphi1}
	\max_{\bPhi} \sum_{k=1}^K\trk_k(\bPhi)-\gamma||\bPhi-V_{RF}(\thetak)||^2\nonumber\\ \mbox{s.t.}\quad
	\la\bPhi^H\bPhi,\sum_{k=1}^{K}[\vBko_k]^2 \ra\leq P,
\end{eqnarray}
with $\trk_k(\bPhi)$ defined from  (\ref{bphi3}). For $\bphi={\sf vec}(\bPhi)$, by  using the identity
$h_k\bPhi\vBko_{\ell}=\hko_{k,\ell}\bphi$
with $\hko_{k,\ell}\triangleq (\vBko_{\ell})^T\otimes h_k\in\mathbb{C}^{1\times (NN_{RF})}$,
we formulate
\[\trk_k(\bPhi)
=\talphak_k+2\Re\{\tak_k\bphi  \}-\la\tXik_k,[\bphi]^2\ra,
\]	
for $\tak_k={\sf vec} [(\tAk_k)^T]^T$ and 	 $\tXik_k\triangleq \tbetak_k\sum_{\ell=1}^K[(\hko_{k,\ell})^H]^2$.
Then
\begin{eqnarray}
&&\sum_{k=1}^K\trk_k(\bPhi)-\gamma||\bPhi-V_{RF}(\thetak)||^2\nonumber\\
&=&\talphak+2\Re\{ \tak\bphi \}-\la\tXik,[\bphi]^2\ra,\nonumber
\end{eqnarray}
for  $\talphak\triangleq \sum_{k=1}^K\talphak_k-\gamma NN_{RF}$ and
$
\tak\triangleq \sum_{k=1}^K\tak_k+\gamma ({\sf vec}(V_{RF}(\thetak)))^H$,  $\tXik\triangleq 	\sum_{k=1}^K\tXik_k+\gamma I_{NN_{RF}}$.
	
Furthermore, we have	$\bPhi\sum_{k=1}^{K}\vBko_k=\clAko\bphi$,
for	$\clAko\triangleq \left(\sum_{k=1}^K\vBko_k\right)^T\otimes I_N$,
so the problem (\ref{sumbbphi1}) is reformulated as
\begin{eqnarray}\label{sumbbphi1e}
	\max_{\bphi} \talphak+2\Re\{ \tak\bphi \}-\bphi^H\tXik\bphi\nonumber\\
	\mbox{s.t.}\quad \la[(\clAko)^H]^2,[\bphi]^2\ra\leq P,
\end{eqnarray}
which admits the closed-form solution of
\begin{equation}\label{sumbbphi5}
	\phiko=\begin{cases}\begin{array}{ll}
			\left(\tXik \right)^{-1}(\tak)^H\\ \mbox{if}\quad ||\clAko\left(\tXik \right)^{-1}(\tak)^H||^2\leq P\cr
			\left(\tXik+\lambda  [(\clAko)^H]^2\right)^{-1}(\tak)^H\\ \mbox{otherwise},
		\end{array}
	\end{cases}
\end{equation}
where $\lambda>0$ is found by bisection, so that  $||\clAko \left(\tXik+\lambda  [(\clAko)^H]^2\right)^{-1}(\tak)^H||^2=P$.
\subsubsection{Algorithm}
Thus in parallel to Algorithm \ref{salg1}, Algorithm \ref{sfalg1} presents a scalable computational procedure for the solution of  (\ref{sumbmm1}), {\color{black} with the total computational complexity of each iteration being on the order of ${\cal O}(N_{RF}K)+{\cal O}(N_{RF}N)$}.

\begin{algorithm}[!t]
	\caption{Scalable SR maximization-based FC HBF algorithm } \label{sfalg1}
	\begin{algorithmic}[1]
		\State \textbf{Initialization:} Initialize $(\theta^{(0)}, \Phi^{(0)}, v^{B,(0)})$.
		\State \textbf{Repeat until convergence of the objective function in  (\ref{sumbmm1}):} Generate  $\vBko$ by
		(\ref{sumsuad6}),
		and $\Phiko$  by  (\ref{sumbbphi5}), and $\thetako$ by (\ref{bthe3}).
		Reset $\kappa:=\kappa+1$.
		\State \textbf{Output} $(\theta^{opt}, \tilde{v}^{B,opt})=(\thetak,\vBk)$.
	\end{algorithmic}
\end{algorithm}
\subsection{Soft max-min rate optimization-based design}
Similarly to (\ref{ssa1}), the soft max-min rate problem is addressed via the following problem of penalized optimization:
\begin{equation}\label{bsmm1}
	\min_{\btheta,\bv^B,\bPhi}\ [\ln\pi_c(\bPhi,\bv^B) +\gamma||\bPhi-V_{RF}(\btheta)||^2]\quad \mbox{s.t.}\quad (\ref{br1}), (\ref{bmm1b}).
\end{equation}
where $\gamma>0$ is a penalty parameter, and
\[
\pi_c(\bPhi,\bv^B)\triangleq \sum_{k=1}^{K}\left(1- \frac{|\hbar_k(\bPhi)\bv^B_k|^2}{|\hbar_k(\bPhi)\bv^B_k|^2+c\psi_k(\bPhi,\bv^B)}\right).
\]
\subsubsection{Alternating optimization in  BBF}
$\vBko$ is generated by solving the following problem:
\begin{equation}\label{bsad9}
	\min_{\bv^B}\ \rhok(\bv^B)\quad\mbox{s.t.}\quad
	\sum_{k=1}^{K} \la  (\Phik)^H\Phik,[\bv^B_k]^2 \ra\leq P,
\end{equation}
where $\rhok(\bv^B)$ is a tight majorant of $\ln \pi_c(\Phik,\bv^B)$ defined by
\[
\rhok(\bv^B)\triangleq\ak+2\sum_{k=1}^K\Re\{\bk_k\bv^B_k\}-\sum_{k=1}^K\la\Ck_k,[\bv^B_k]^2\ra
\]
with
\[
\begin{array}{c}\ak\triangleq \ln\pi_c(\Phik,\vBk)+\sum_{k=1}^K\dk_k|\bar{x}_k|^2+c\sigma\sum_{k=1}^K\ck_k,\\
	\bk_k\triangleq \dk_k (\hbar_k(\Phik)\vBk_k)^*\hbar_k(\Phik),\\
	\Ck_k\triangleq\ds c\sum_{\ell\in\clK\setminus\{k\}}\ck_{\ell}[\hbar^H_{\ell}(\Phik)]^2+
	\ck_k[\hbar^H_{k}(\Phik)]^2,
\end{array}\]		
for $\dk_k\triangleq
\frac{(c\bar{y}_k+|\bar{x}_k|^2)^{-1}}{\pi_c(\Phik,\vBk)}$, $\ck_k\triangleq \dk_k\frac{|\bar{x}_k|^2}{c\bar{y}_k+|\bar{x}_k|^2}$, $\bar{x}_k\triangleq \hbar_k(\Phik)\vBk$, and
$\bar{y}_k\triangleq \psi_k(\Phik,\vBk)$.
The problem (\ref{bsad9}) admits
the closed-form solution of
\begin{equation}\label{bsad10}
	\vBko_k=\begin{cases}\begin{array}{ll}(\Ck_k)^{-1}(\bk_k)^H\\
			\mbox{if}\quad \ds\sum_{k=1}^K \la [(\Phik)^H]^2,[(\Ck_k)^{-1}(\bk_k)^H]^2\ra\leq P,\cr
			(\Ck_k+\lambda [(\Phik)^H]^2 )^{-1}(\bk_k)^H\\ \mbox{otherwise},
		\end{array}
	\end{cases}
\end{equation}
where $\lambda>0$ is found by bisection, so that
$	\sum_{k=1}^K\la [(\Phik)^H]^2,[(\Ck_k+\lambda [(\Phik)^H]^2 )^{-1}(\bk_k)^H]^2\ra=P$.
\subsubsection{Alternating optimization in $\bPhi$}
$\phiko\triangleq {\sf vect}(\Phiko)$ is generated by solving the problem
\begin{eqnarray}\label{bsphi2m}
	\max_{\bphi\triangleq {\sf vect}(\bPhi)} \trhok(\bphi)-\gamma ||\bphi-v_{RF}(\thetak)||^2\nonumber\\
	\mbox{s.t.}\quad ||\clAko\bphi||^2\leq P,
\end{eqnarray}
where $\trhok(\bphi)$ is a tight majorant of $\ln\pi_c(\bPhi,\vBko)$ defined by
\[
\trhok(\bphi)\triangleq \tak+2\Re\{\tbk\bphi\}-\bphi^H\tCk\bphi
\] with
\[
\begin{array}{c}
	\tak\triangleq \begin{aligned}[t]&\ln\pi_c(\Phik,\vBko)	-\sum_{k=1}^K\tdk_k|\bar{x}_k|^2\\&-c\sigma\sum_{k=1}^K\tck_k,\end{aligned}\\
	\tbk\triangleq \sum_{k=1}^K\left( \tdk_k (\hko_{k,k}\varphik)^*\hko_{k,k}\right),\\
	\tCk\triangleq\ds \sum_{k=1}^K\tck_{k}\left( c\sum_{\ell\in\clK\setminus\{k\}}[(\hko_{k,\ell})^H]^2+
	[(\hko_{k,k})^H]^2\right),
\end{array}
\]		
for $\tdk_k\triangleq \frac{(c\bar{y}_k+|\bar{x}_k|^2)^{-1}}{\pi_c(\Phik,\vBko)}$, $\tck_k\triangleq\tdk_k\frac{|\bar{x}_k|^2}{c\bar{y}_k+|\bar{x}_k|^2}$, and
$\bar{x}_k=\hbar_k(\Phik)\vBko_k$, $\bar{y}_k=\psi_k(\Phik,\vBko)$.

The problem (\ref{bsphi2m}) admits the closed-form solution of
\begin{equation}\label{bssad10}
	\phiko=\begin{cases}\begin{array}{l}(\tCk+\gamma I_{NN_{RF}})^{-1}((\bk)^H+\gamma v_{RF}(\thetak))\cr
			\mbox{if}\quad \ds\sum_{k=1}^K ||\clAko(\tCk+\gamma I_{NN_{RF}})^{-1}\\\quad\quad ((\bk)^H+\gamma v_{RF}(\thetak))||^2 \leq P,\cr
			(\Ck_k+\gamma I_{NN_{RF}}+\lambda  [(\clAko)^H]^2)^{-1}\\((\bk)^H+\gamma v_{RF}(\thetak))\\ \mbox{otherwise},
		\end{array}
	\end{cases}
\end{equation}
where $\lambda>0$ is found by bisection, so that
$
||\clAko(\Ck_k+\gamma I_{NN_{RF}}+\lambda [(\clAko)^H]^2)^{-1}((\bk)^H+\gamma v_{RF}(\thetak))||^2=P$.
\subsubsection{Alternating optimization in ABF}
Generate  $\thetako$ according (\ref{the3}).
\subsubsection{Algorithm}
Thus in parallel to Algorithm \ref{alg2}, Algorithm \ref{falg2} presents a scalable computational procedure for the solution of  (\ref{bsmm1}), {\color{black} with the total computational complexity of each iteration being the same as that of Algorithm \ref{sfalg1}}.
\begin{algorithm}[!t]
	\caption{Scalable soft max-min optimization-based FC HBF algorithm } \label{falg2}
	\begin{algorithmic}[1]
		\State \textbf{Initialization:} Initialize $(\theta^{(0)}, \Phi^{(0)}, v^{B,(0)})$.
		\State \textbf{Repeat until convergence of the objective function in  (\ref{bsmm1}):} Generate  $\vBko$ by (\ref{bsad10}),
		$\Phiko$  by (\ref{bssad10}), and $\thetako$ by (\ref{the3}).
		Reset $\kappa:=\kappa+1$.
		\State \textbf{Output} $(\theta^{opt}, v^{B,opt})=(\thetak,\vBk)$.
	\end{algorithmic}
\end{algorithm}
\section{Numerical Results}\label{sec:simulation}
This section analyzes the performance of the proposed algorithms along with their computational convergence.
The number of downlink transmit antennas (DL-TA) at the BS is $N = 72$, and that of UEs is $K = 8$. All the UEs are randomly placed in a cell of radius $200$ meters.
The path-loss of UE $k$ experienced at a distance $d_k$ from the BS is set to $\rho_k = 36.72 + 35.3 \log10(d_k) $ (in dB), taking into account a $16.5$ dB gain provided by multiple-antenna aided mmWave transmission \cite{GPP10,Akd14,Rappa17}.
The  channel $h_k\in\mathbb{C}^{1\times N}$  between  the  BS and  UE $k$  is modelled by \cite{Ayetal14} $h_k=
F \sqrt{10^{-\rho_k/10}} \sum_{c=1}^{N_c} \sum_{\ell = 1}^{N_{sc}} \alpha_{k,c,\ell}  a \left(\phi_{k,c,\ell},\theta_{k,c,\ell} \right)$,
where $F =  \sqrt{\frac{N}{N_c N_{sc}}}$, $N_c$ is the number of scattered clusters, $N_{sc}$ is the number of scatterers within each cluster, and $\alpha_{k,c,\ell} \sim \mathcal{CN}(0,1)$ is the complex gain of the $\ell$th path in the $c$th cluster between the BS and UE $k$. We set $N_c = 5$ and $N_{sc} =10$ as in \cite{Akd14}. Assuming a uniform planar antenna array configuration having half-wavelength antenna spacing with $N_1$ and $N_2$ elements in the horizontal and vertical dimensions, respectively, the normalized antenna array response vectors  $a \left(\phi_{k,c,\ell},\theta_{k,c,\ell} \right)$ is defined as
\[
\begin{array}{r}
	\begin{aligned}[t]&a \left(\phi_{k,c,\ell},\theta_{k,c,\ell} \right)\\=&
	\frac{1}{\sqrt{N}} \left( 1, e^{j \pi ( x \sin( \phi_{k,c,\ell}^t )\sin(\theta_{k,c,\ell}^t) + y \cos(\theta_{k,c,\ell}^t)) },\nonumber \hdots,\right.\\
	&\left. e^{j \pi ((N_1 - 1 )\sin( \phi_{k,c,\ell}^t )\sin(\theta_{k,c,\ell}^t) + (N_2  - 1)\cos(\theta_{k,c,\ell}^t)) }  \right)^T,\end{aligned}
\end{array}\]
where $0 \leq x \leq (N_1 - 1) $ and $0 \leq y \leq (N_2 - 1)$,  $\phi_{k,c,\ell}$   and $\theta_{k,c,\ell}$  are the azimuth angle and elevation angle of departure for the $\ell$th path in the $c$th cluster arriving from the BS to the UE $k$, respectively. The angles are generated using the Laplacian distribution in combination with random mean cluster angles in the interval of $[0, 2\pi)$ and a 10-degree spread for each cluster. Assuming a carrier frequency of 28 GHz, the noise power density is $-$174 dBm/Hz. The results are multiplied by $\log_2 e$ to convert the unit nats/sec into the unit bps/Hz.  The algorithm terminates, when the penalty term falls below $10^{-2}$.

We use the following legends to specify the proposed implementations:
\begin{itemize}
	\item For the AOSA based algorithms, ``max-min'' and ``3-bit max-min'' refer to the convex-solver-based Algorithm \ref{alg1} with the ABF matrix having $\infty$ resolution and 3-bit resolution, respectively;  ``soft max-min'' and ``3-bit soft max-min'' refer to the scalable Algorithm \ref{alg2} with the ABF matrix having $\infty$ resolution and 3-bit resolution, respectively;  ``SR'' and ``3-bit SR'' refer to Algorithm \ref{salg1} with the ABF matrix having $\infty$ resolution and 3-bit resolution, respectively.
	
	\item For the FC HBF algorithms, ``FC max-min'' refers to Algorithm \ref{falg1} with the ABF matrix having $\infty$ resolution; ``FC soft max-min'' refers to Algorithm \ref{falg2} with the ABF matrix having $\infty$ resolution; ``FC SR'' refers to Algorithm \ref{sfalg1} with the ABF matrix having $\infty$ resolution; {\color{black}``FC RZFB max-min'' refers to the regularized zero-forcing beamforming (RZFB)-aided max-min rate algorithm proposed in \cite{Yuetal23twc}. The superior performance of FC RZFB over  the algorithms in \cite{SY16} and \cite{Nasetal20TVT}
 has been demonstrated by \cite[Fig. 1]{Yuetal23twc} and \cite[Figs. $3$, $6$, $8$]{Yuetal23twc}, respectively. Moreover, the superior performance of the algorithm in \cite{Nasetal20TVT} over
other existing algorithms \cite{SY16,Paetal17,Noh-16-IET-A,Haetal17}  has been demonstrated by
\cite[Fig. 14]{Nasetal20TVT}. Thus FC RZFB max-min serves as the baseline algorithm, since it outperforms the existing algorithms \cite{SY16,Paetal17,Noh-16-IET-A,Haetal17};\footnote{\color{black}It should be noted that in this paper we only consider  cases of low
$N_{RF}\leq K$, for which the zero-forcing beamforming-aided algorithm proposed in \cite{Cui20wcl}   is not applicable.}
``FC Shi-Hong'' refers to the SR algorithm proposed in \cite{Shi-18-Jun-A}; ``FC Zhang et al.'' refers to the algorithm proposed in \cite{Zhaetal20a,Zhaetal20tvt}, which only works for the case of $N_{RF}=K$.}
\end{itemize}
\subsection{AOSA vs. FC}

\begin{table*}[t]
	\centering
	\caption{The SR achieved upon maximizing it at $P_{ref} = 38.02$ dBm}
	\begin{tabular}{|l|c|c|c|c|}
		\hline
		& SR ($N_{RF}=8$) & SR ($N_{RF}=6$) & SR ($N_{RF}=4$) & FC SR ($N_{RF}=4$) \\ \hline
		Achieved SR (bps/Hz) & 63.9    & 51.4     & 35.8    & 33.2  \\ \hline
	\end{tabular}
	\label{table:SR_AOSA_FC}
\end{table*}

\begin{table*}[t]
	\centering
	\caption{The average number of ZRs in maximizing SR at $P_{ref} = 38.02$ dBm}
	\begin{tabular}{|l|c|c|c|c|}
		\hline
		& SR ($N_{RF}=8$) & SR ($N_{RF}=6$) & SR ($N_{RF}=4$) & FC SR ($N_{RF}=4$) \\ \hline
		The average \# of ZRs & 0.2    & 2.0     & 4.0    & 4.0  \\ \hline
	\end{tabular}
	\label{table:SR_num_low_rate_FC}
\end{table*}

\begin{figure}[!t]
	\centering
	\includegraphics[width=0.85\linewidth]{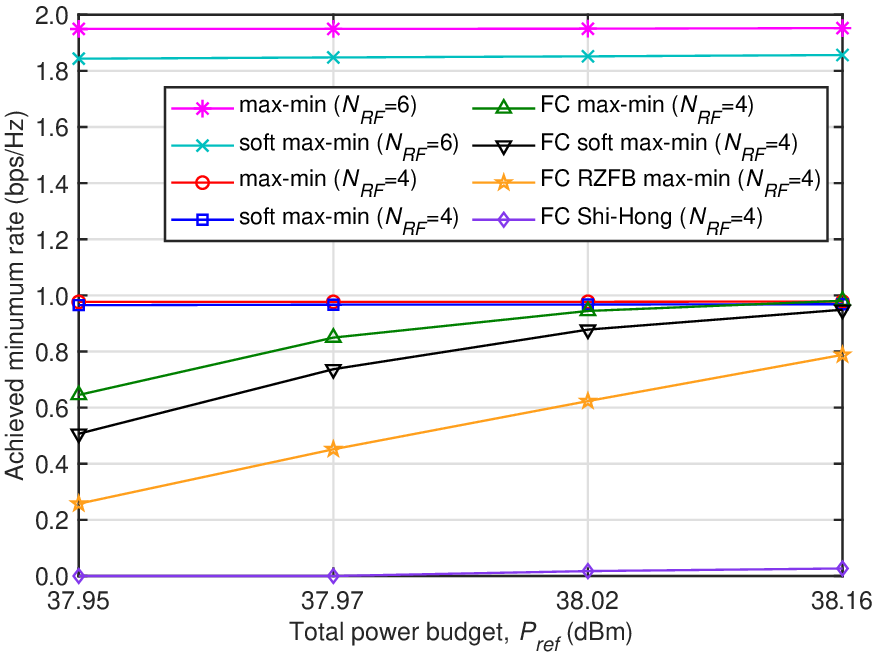}
	\caption{The achievable MR vs. the total power $P_{ref}$.}
	\label{fig:FC_MR_min_rate_vs_Pref}
\end{figure}

\begin{figure}[!t]
	\centering
	\includegraphics[width=0.85\linewidth]{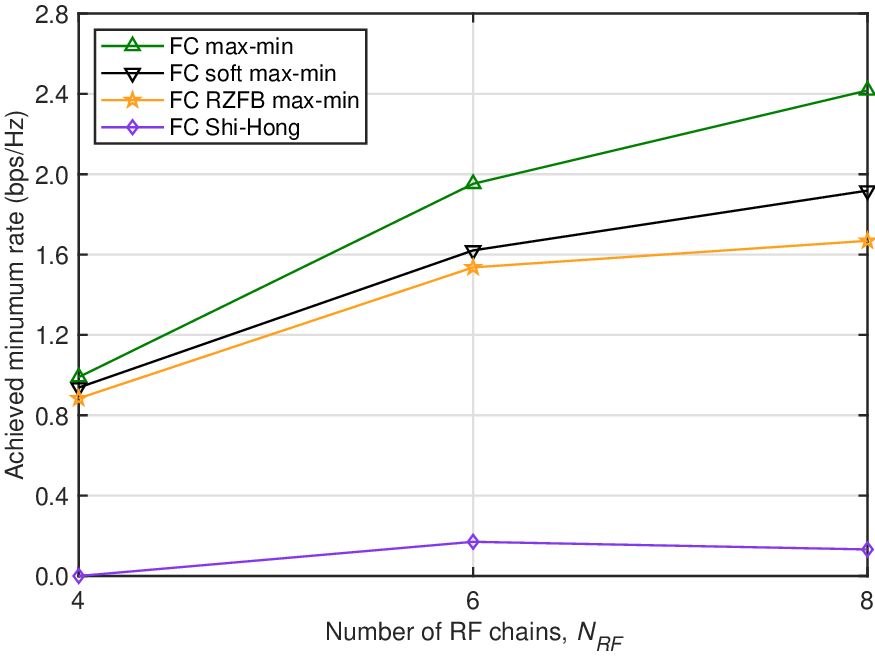}
	\caption{MR of max-min and soft max-min vs. existing algorithms under different number of RF chains $N_{RF}$.}
	\label{fig:FC_RZF_MR_min_rate_vs_RF}
\end{figure}

\begin{figure}[!t]
	\centering
	\includegraphics[width=0.85\linewidth]{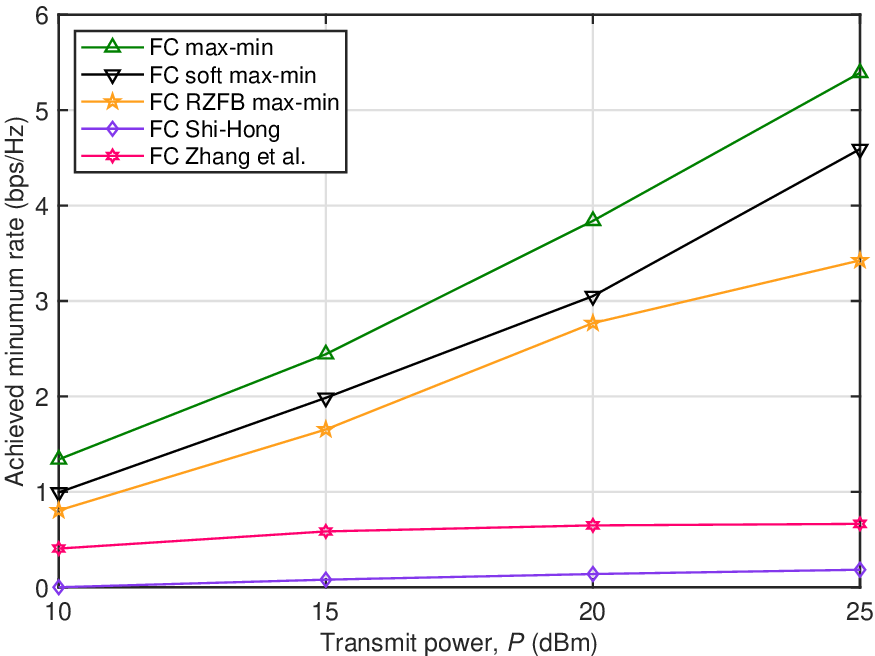}
	\caption{MR of max-min and soft max-min vs. existing algorithms under $N_{RF} = K$ and different transmit power $P$.}
	\label{fig:FC_alg_comp_min_rate_vs_P}
\end{figure}

We start by evaluating the performance of the AOSA and FC structures.
To ensure a fair comparison, we use the total power in the case of FC as the reference, which is defined as
\begin{equation}\label{pref}
	P_{ref}=\bar{P}+ \bar{N}_{RF}\times 118 +N\bar{N}_{RF}\times 20,
\end{equation}
where $\bar{P}$ is the transmit power in mW, $\bar{N}_{RF}$ is the original number of RF chains used for FC, $118$ mW is the circuit power consumption  per  RF chain, and $20$ mW is the circuit power consumption per  phase shifter \cite{LL16}. To determine the total power budget $P_{ref}$, we fix $\bar{N}_{RF}$ to $4$ and vary the transmit power $\bar{P}$ from $10$ dBm ($10$ mW) to $25$ dBm ($316$ mW) at $5$ dBm ($3$ mW) intervals. Then the relationship between the transmit power $P$ and $N_{RF}$ for the AOSA structure is
\begin{equation}\label{ref}
	P+N_{RF}\times 118+ N\times 20 = P_{ref}.
\end{equation}

Table \ref{table:SR_AOSA_FC} compares the SR maximized by the SR maximization-based Algorithms 2 and 5 at  $P_{ref} = 38.02$ dBm ($\bar{P}=20$ dBm). The AOSA  achieves a higher SR than the FC, because the former enables us to allocate much more transmit power and exploit more RF chains, hence resulting in more effective spatial DL beamforming under the same total power budget.

We define the negligible rate of less than $0.001$ bps/Hz  as zero rate (ZR).
Table \ref{table:SR_num_low_rate_FC} displays the number of  ZR users under maximizing the SR. The results  demonstrate that SR maximization is not suitable for  MU services, although increasing the number of RF chains for transmitting more data streams also goes some way towards reducing
the average number of ZR UEs. Moreover, the average number of ZR UEs is 0.2 when $N_{RF}=8$, indicating that increasing the number of RF chains mitigates the problem prevent them in most cases. However, the transmit powers $P$ required by the AOSA-based SR algorithms with $N_{RF}=4$, $N_{RF}=6$, and $N_{RF}=8$ are 36.45 dBm, 36.22 dBm, and 35.96 dBm, respectively, which are impractically high.

{\color{black}In Fig. \ref{fig:FC_MR_min_rate_vs_Pref}, we present a performance comparison between the two structures using our max-min-based algorithms and the existing algorithms of \cite{Yuetal23twc,Shi-18-Jun-A}.}  AOSA outperforms FC in terms of the achieved MR and the performance gap becomes wider, when more RF chains are utilized. Furthermore, our proposed soft max-min algorithm is capable of achieving MR that is comparable to those obtained by the convex-solver-based max-min algorithm.  It follows from (\ref{pref}) and (\ref{ref}) that under the same $P_{ref}$, the transmit power $P$ of
AOSA in (\ref{ref}) is very high compared to that of its FC counterpart $\bar{P}$ in (\ref{pref}). The former is not sensitive to the value of the latter in the interval of  $10$ dBm to $25$ dBm. This is why the performance of AOSA  is seen to be flat in Fig. \ref{fig:FC_MR_min_rate_vs_Pref}. {\color{black}Additionally, both the FC RZFB max-min and FC Shi-Hong algorithms were implemented under the FC structure associated with $N_{RF} = 4$, as dictated by the specific total power budget $P_{ref}$, and they were also characterized in Fig. \ref{fig:FC_MR_min_rate_vs_Pref}. We can observe that both the FC RZFB max-min and FC Shi-Hong algorithms are outperformed by our proposed max-min and soft max-min algorithms in terms of their MR.}

{\color{black}Fig. \ref{fig:FC_RZF_MR_min_rate_vs_RF} facilitates a comprehensive analysis by comparing our proposed max-min and soft max-min algorithms to the FC RZFB max-min and FC Shi-Hong algorithms, under a fixed total transmit power $P_{ref}$ of $40.97$ dBm upon varying the number of RF chains. In this context, $P_{ref}$ is calculated by setting the number of RF chains to $8$ and $P$ to 15 dBm. Notably, both the max-min and soft max-min algorithms outperform the FC RZFB max-min algorithm for  all the values of $N_{RF}$ considered. Furthermore, they exhibit a significant performance advantage over the SR maximization-based FC Shi-Hong algorithm.}

{\color{black}In Fig. \ref{fig:FC_alg_comp_min_rate_vs_P}, we compare our proposed max-min and soft max-min algorithms to the existing algorithms of \cite{Yuetal23twc,Shi-18-Jun-A,Zhaetal20a,Zhaetal20tvt} while varying the transmit power $P$. To simulate the FC Zhang algorithm \cite{Zhaetal20a,Zhaetal20tvt}, we set $N_{RF} = K = 8$. It is observed that our proposed max-min and soft max-min algorithms exhibit superior performance compared to the others. This is because the FC Zhang algorithm focuses on enhancing the signal energy, rather than effectively mitigating the multi-user interference, hence resulting in a lower minimum rate.}

Hence, from now on, we will utilize AOSA instead of the FC structure. Additionally, we will focus on the max-min-based algorithms because of the deficiency of the ZR SR maximization based algorithms.

\subsection{The AOSA performance under low-resolution ABF}

\begin{table*}[!t]
\centering
\caption{The average number of ZRs in maximizing SR with moderate power allocation}
\begin{tabular}{|l|c|c|c|}
	\hline
	& SR ($N_{RF}=8$) & SR ($N_{RF}=6$) & SR ($N_{RF}=4$) \\ \hline
	The average \# of ZRs & 1.6    & 2.1     & 4.0  \\ \hline
\end{tabular}
\label{table:SR_num_low_rate_AOSA}
\end{table*}

\begin{figure}[!t]
	\centering
	\includegraphics[width=0.85\linewidth]{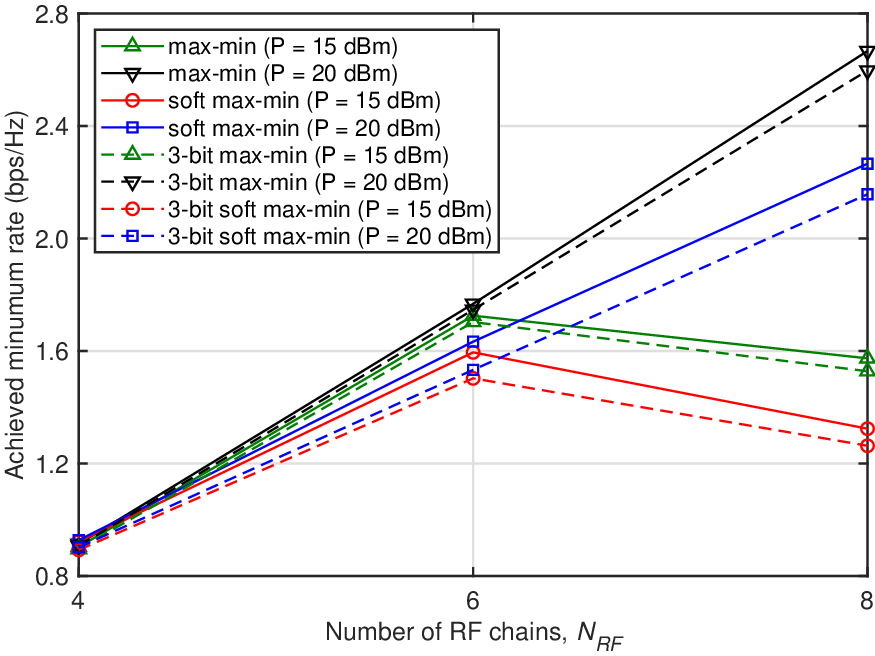}
	\caption{The achieved MR vs. the number of RF chains $N_{RF}$.}
	\label{fig:AOSA_MR_min_rate_vs_RF}
\end{figure}

\begin{figure}[!t]
	\centering
	\includegraphics[width=0.85\linewidth]{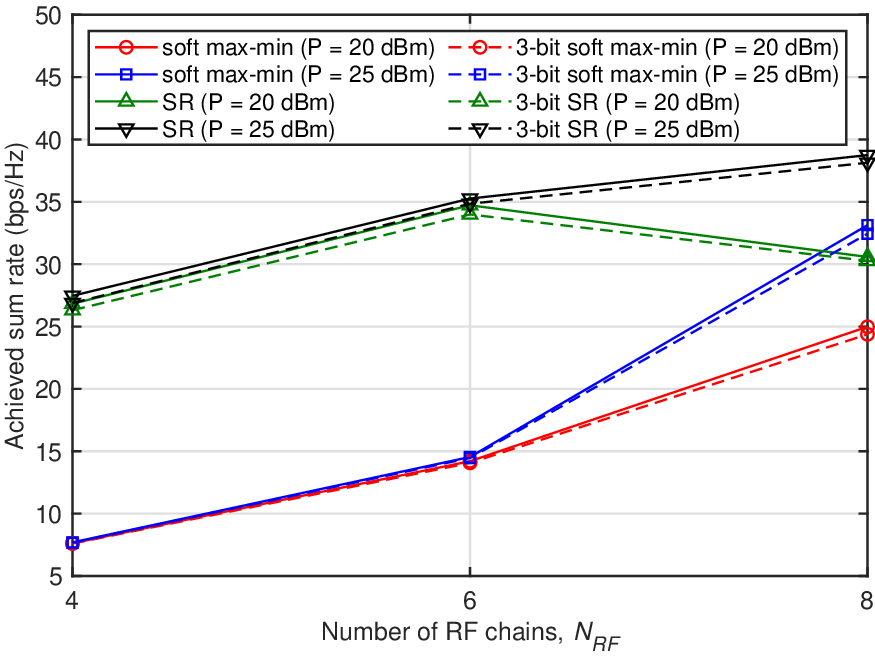}
	\caption{The achieved SR vs. the number of RF chains $N_{RF}$.}
	\label{fig:AOSA_MR_SR_sum_rate_vs_RF}
\end{figure}

\begin{figure}[!t]
	\centering
	\includegraphics[width=0.85\linewidth]{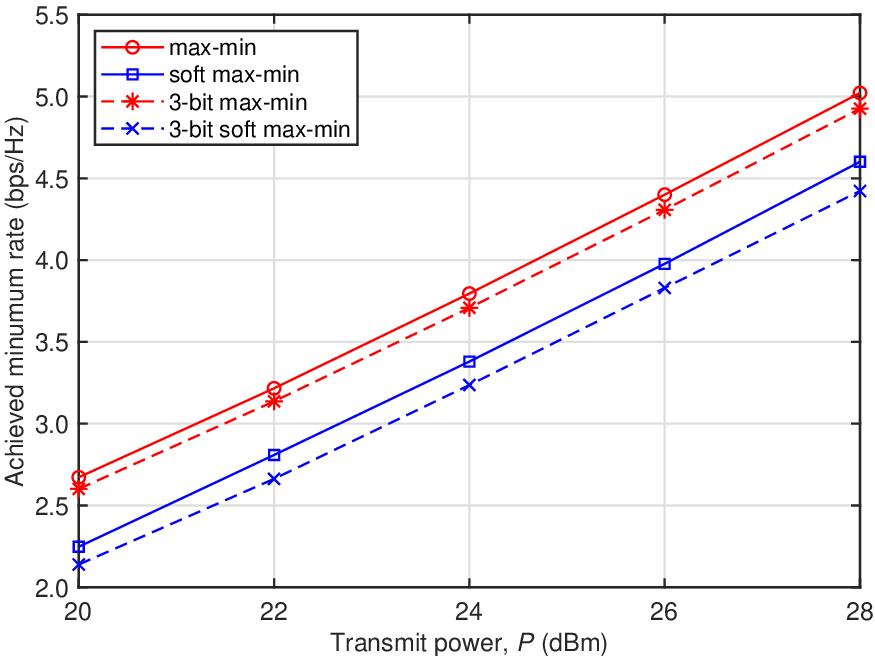}
	\caption{The achieved MR vs. the transmit power $P$.}
	\label{fig:AOSA_MR_min_rate_vs_P}
\end{figure}

\begin{figure}[!t]
	\centering
	\includegraphics[width=0.85\linewidth]{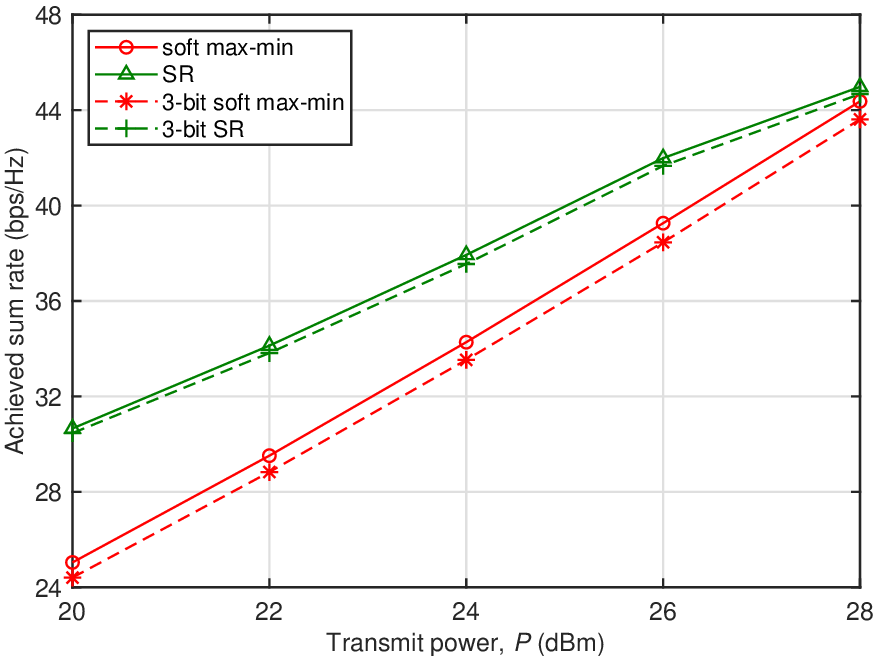}
	\caption{The achieved SR vs. the transmit power $P$.}
	\label{fig:AOSA_MR_SR_sum_rate_vs_P}
\end{figure}

\begin{figure}[!t]
	\centering
	\includegraphics[width=0.85\linewidth]{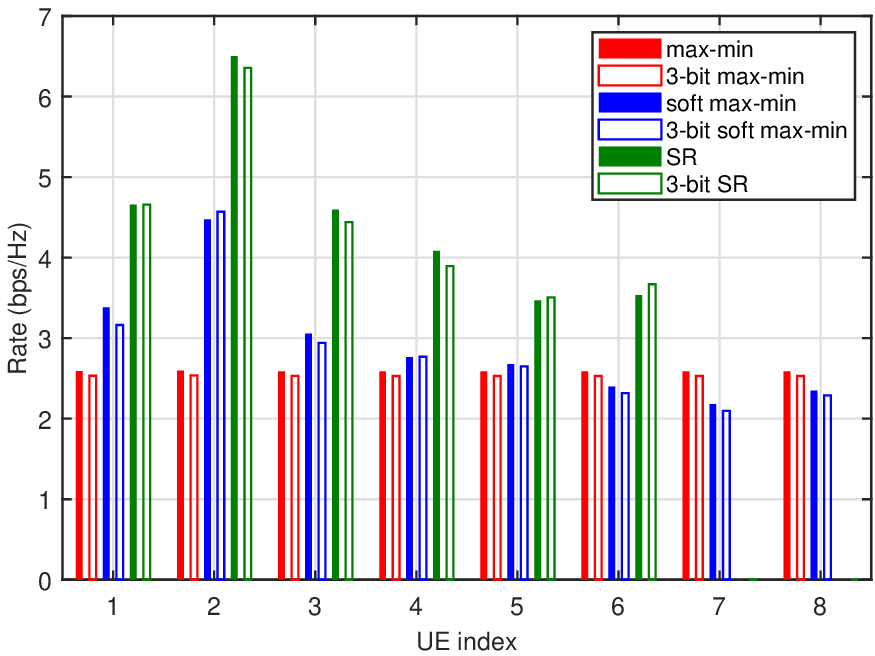}
	\caption{The distribution of user-rates at $N_{RF} = 8$ and $P = 20$ dBm.}
	\label{fig:AOSA_MR_user_rate_P20}
\end{figure}

\begin{table}[!t]
	\centering
	\caption{Min-rate/max-rate and Jains fairness index of user-rate allocation with $N_{RF} = 8$ and $P = 20$ dBm}
	\begin{tabular}{|l|c|c|}
		\hline
		& soft max-min & 3-bit soft max-min \\ \hline
		Min-rate/max-rate     & 0.46       & 0.45             \\ \hline
		Jain's fairness index & 0.91       & 0.90             \\ \hline
	\end{tabular}
	\label{table:fainess_soft_max_min}
\end{table}

\begin{figure}[!t]
	\centering
	\includegraphics[width=0.85\linewidth]{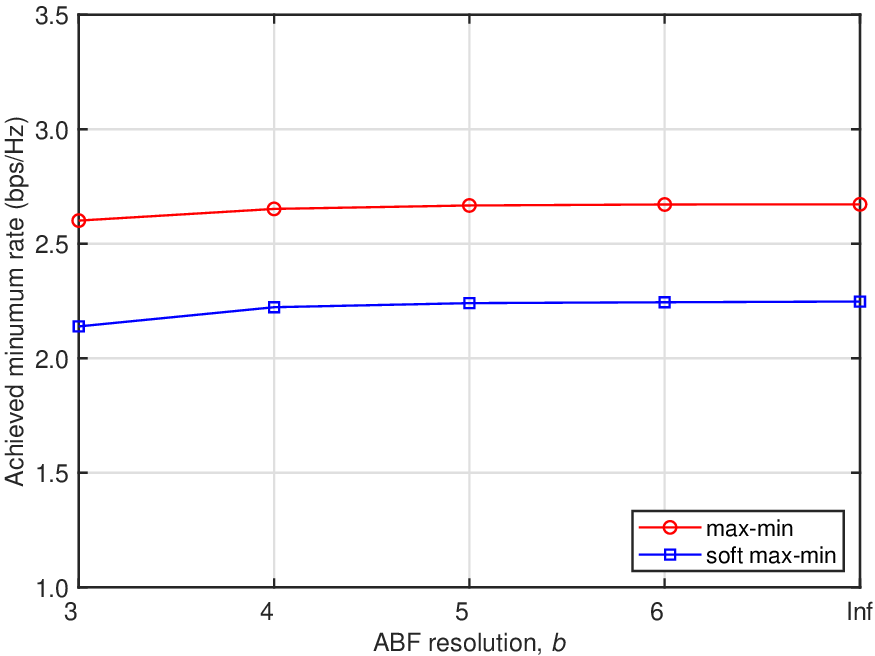}
	\caption{The achieved MR vs. ABF resolution $b$.}
	\label{fig:AOSA_MR_min_rate_vs_bit}
\end{figure}

In this subsection, we initially evaluate the performance by varying the numbers of RF chains. To ensure a fair comparison, we maintain a constant total power budget $P_{ref}$, while increasing the number of RF chains. This results in additional ``virtual antennas" for increased angular flexibility, but at a reduced transmit power budget $P$. Thus, we have to find a balance between the number of RF chains $N_{RF}$ and the transmit power budget $P$.
For calculating the total reference power $P_{ref}$, we set the number of RF chains to $8$ and $P$ to 15 dBm. The resultant total power budget is $P_{ref} = 33.83$ dBm.
Fig. \ref{fig:AOSA_MR_min_rate_vs_RF} shows the MR achieved by the max-min-based algorithms.
When we reduce the number of RF chains from 8 to 6, more transmit power is allocated at the baseband, which results in a higher MR achieved by $N_{RF} = 6$ than by $N_{RF} = 8$. However, despite the higher transmit power budget at $N_{RF}=4$, using 8 RF chains still outperforms using 4 RF chains, demonstrating the advantages of utilizing more RF chains for enhancing the digital beamforming part.
We then increase the transmit power $P$ to investigate the conditions under which digital beamforming using 8 RF chains can be fully leveraged. At $P = 20$ dBm, where $P_{ref} = 33.95$ dBm, the achievable MR associated with $N_{RF} = 8$ substantially outperforms the configurations using fewer RF chains but higher transmit power allocation.
Essentially, when the total power budget is limited, we opt for utilizing a smaller number of RF chains to allow for having an adequate transmission power. However, to fully exploit digital beamforming utilizing a larger number of RF chains, a higher transmit power is necessary.

Then in Fig. \ref{fig:AOSA_MR_SR_sum_rate_vs_RF}, we compare the SRs obtained by the soft max-min algorithm and the SR maximization-based algorithm. It can be observed that increasing the number of RF chains and the BS transmit power also leads to an increase in SR for the soft max-min algorithm. However, the SR maximization-based algorithm requires a higher transmit power $P$ to fully exploit the benefits of utilizing 8 RF chains. As the transmit power budget $P$ is increased to 25 dBm, the benefits of 8 RF chains become more substantial than those of 6 RF chains for the SR maximization-based algorithm. It is noteworthy that when the transmit power budget $P$ is set to 25 dBm and $N_{RF}$ is increased from 6 to 8, the performance gap between the soft max-min algorithm and the SR maximization-based algorithm narrows significantly. This inspires us to improve the performance of the soft max-min algorithm in terms of its SR by utilizing a larger number of RF chains and a higher transmit power budget, allowing it to match the the performance of the SR maximization-based algorithm.
Additionally, Fig. \ref{fig:AOSA_MR_min_rate_vs_RF} and Fig. \ref{fig:AOSA_MR_SR_sum_rate_vs_RF} illustrate that the performance of MR and SR using 3-bit resolution is comparable to that of the $\infty$ resolution scheme, which is also evident from Fig. \ref{fig:AOSA_MR_min_rate_vs_bit}.

Furthermore, we present Table \ref{table:SR_num_low_rate_AOSA} to summarize the average number of ZR UEs resulting from maximizing the SR under moderate BS transmit power in Fig. \ref{fig:AOSA_MR_SR_sum_rate_vs_RF}. Compared  to the observation made concerning Table \ref{table:SR_num_low_rate_FC}, we can see that when the transmit power $P$ of the SR algorithms using $N_{RF}=4$, $N_{RF}=6$, and $N_{RF}=8$ reduced to 27.57 dBm, 25.26 dBm, and 20 dBm, respectively, maximizing the SR using $N_{RF}=8$ still results in low-rate connections or even in ZR UEs in energy-efficient signal transmission scenarios.

Fig. \ref{fig:AOSA_MR_min_rate_vs_P} depicts the MR achieved by the max-min-based algorithms as the transmit power $P$ at $N_{RF} = 8$. Similar to the observation in Fig. \ref{fig:AOSA_MR_min_rate_vs_RF}, the MR achieved using 3-bit resolution is comparable to that of the $\infty$ resolution, and the soft max-min algorithm achieves a slightly lower MR than to the max-min algorithm. In Fig. \ref{fig:AOSA_MR_SR_sum_rate_vs_P}, we compare the SR obtained by the max-min-based algorithms to that of the SR maximization-based Algorithm \ref{salg1} for $N_{RF} = 8$. The soft max-min algorithm achieves much higher SR than the max-min algorithm. It is worth noting that as the transmit power $P$ increases, the SR obtained by the soft max-min algorithm approaches that of the SR maximization-based algorithm. In other words, we can approach the optimal MR and SR by the proposed soft max-min algorithm. This is a surprise, because it is commonly maintained that the MR and SR performances are conflicting, i.e. one of them must be sacrificed to improve the other.

To demonstrate the ability of our proposed soft max-min algorithm to achieve a fair rate allocation, Fig. \ref{fig:AOSA_MR_user_rate_P20} portrays the user-rate distribution pattern obtained for $N_{RF} = 8$ and $P = 20$ dBm. We can observe that the soft max-min algorithm achieves a MR that is comparable to that of the max-min algorithm, while maintaining a good SR. By contrast, maximizing the SR results in the allocation of ZR, thereby literally disconnecting certain UEs. To provide a more detailed analysis, Table \ref{table:fainess_soft_max_min} quantifies the fairness of the user-rate distribution by evaluating both the ratio of min-rate to max-rate and Jain’s fairness index of user-rate allocation \cite{Jain84}  for $N_{RF} = 8$ and $P = 20$ dBm. The results show that the soft max-min algorithm yields a Jain's fairness index that is closer to one, indicating that it achieves a distribution of user-rates that is nearly uniform.

Fig. \ref{fig:AOSA_MR_min_rate_vs_bit} shows the MR achieved for different resolutions of the ABF, given $N_{RF} = 8$ and $P = 20$ dBm. The performance achieved using 5-bit and 6-bit resolutions is nearly indistinguishable from that of the $\infty$ resolution case. However, for 3-bit resolution, the max-min algorithm shows an approximate reduction of $3$\% compared to the $\infty$ resolution, while the soft max-min algorithm shows an approximate reduction of $5$\% compared to the $\infty$ resolution.

\subsection{Algorithmic convergence}
\begin{figure}[t]
	\centering
	\includegraphics[width=0.85\linewidth]{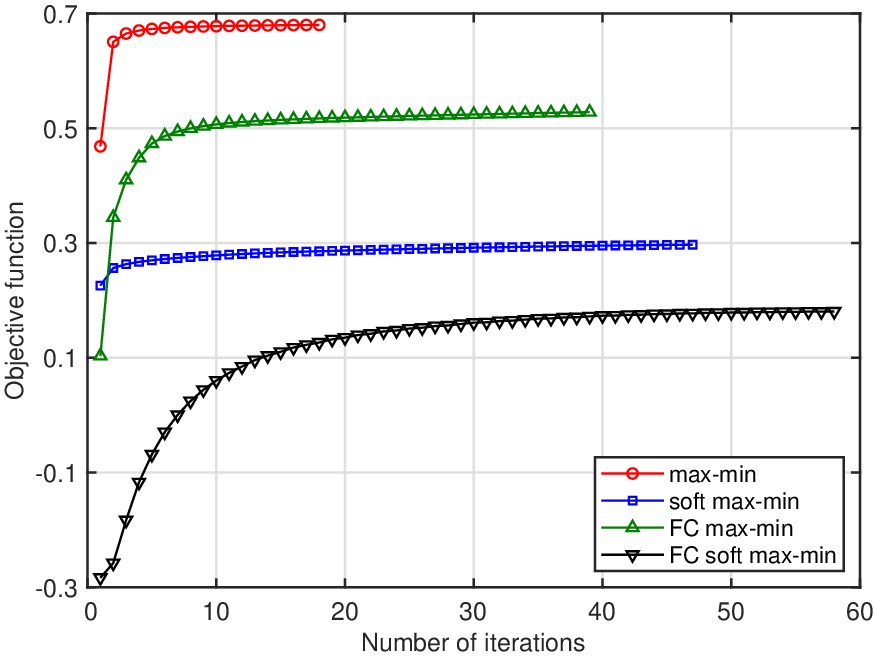}
	\caption{Convergence of the algorithms}
	\label{fig:conv_obj_fun_RF4}
\end{figure}

\begin{figure}[t]
	\centering
	\includegraphics[width=0.85\linewidth]{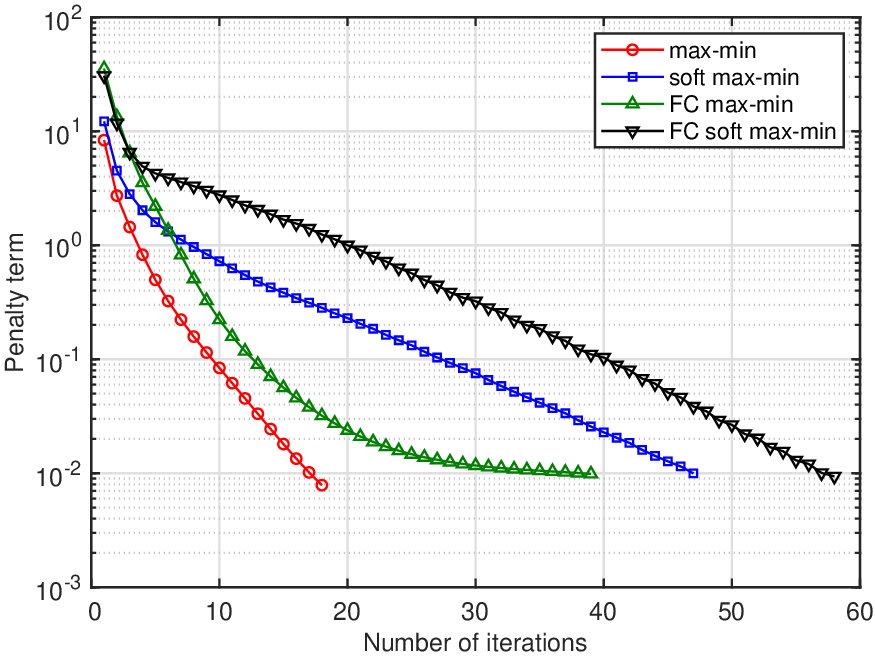}
	\caption{Convergence of penalty terms}
	\label{fig:conv_penalty_RF4}
\end{figure}

\begin{figure}[t]
	\centering
	\includegraphics[width=0.85\linewidth]{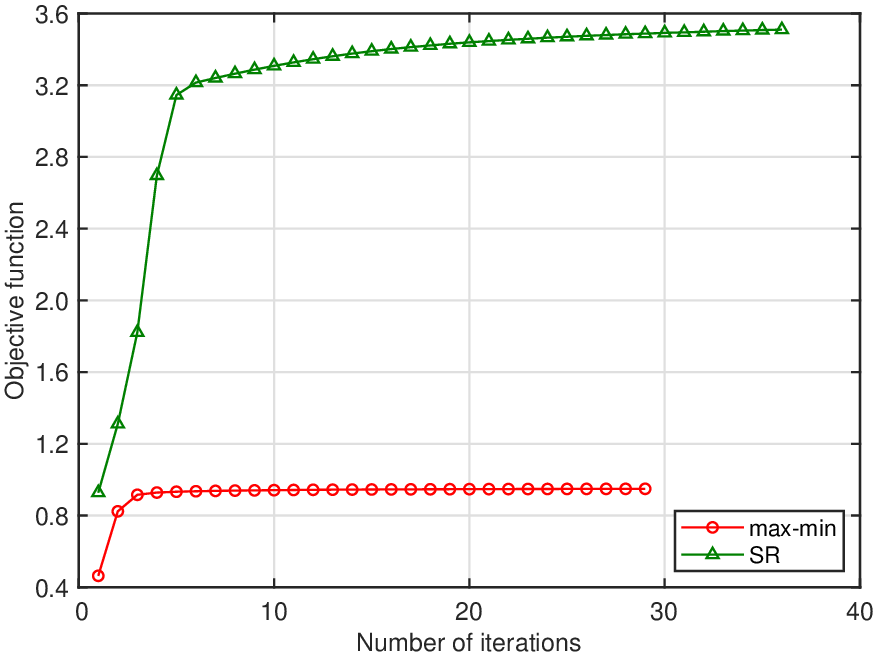}
	\caption{Convergence of the objective functions in  (13) and (31)}
	\label{fig:conv_obj_fun_no_penalty_RF4}
\end{figure}

\begin{table}[!t]
	\centering
	\caption{The achieved MR vs. $c$.}
	\begin{tabular}{|l|c|c|c|}
		\hline
		& $c=1$  & $c=0.1$ & $c=0.01$ \\ \hline
		soft max-min       & 2.1118 & 2.2351  & 1.5217   \\ \hline
		3-bit soft max-min & 2.0440 & 2.1388  & 1.4468   \\ \hline
	\end{tabular}
	\label{table:AOSA_min_rate_vs_c}
\end{table}

Finally, we characterize the convergence of the proposed algorithms.
In our simulations, the coefficient $c$ used in soft max-min based algorithms is set to 0.1. It should be noted that the setting of $c$ is not constant and should be selected appropriately based on the specific scenario.
{\color{black} We present Table \ref{table:AOSA_min_rate_vs_c} to illustrate the impact of $c$ on the MR achieved. The table reveals that the highest MR is attained when $c= 0.1$.}
{\color{black}To ensure a reasonable convergence speed  with the penalty parameter $\gamma$, we begin by selecting an initial value  for $\gamma$ so that the penalty term's magnitude is comparable to that of the objective. As the iterations progress, we gradually increase the value of $\gamma$.
For instance, let's consider the penalty parameter $\gamma$ for implementing
Algorithm \ref{alg1}. We randomly generate $\phi^{(0)}$ with the modulus of its entries being less than $1$ and $v^{B, (0)}$ satisfying the power constraint  (\ref{sa6}). Then
the triplet $(\theta^{(0)},\phi^{(0)},v^{B,(0)})$ with $\theta^{(0)}_{\ell,j}=\lfloor\angle\phi^{(0)}_{\ell,j}\rceil_{b}$, $(\ell,j)\in\clL\times\clN_{RF}$
(see (\ref{the3}))
is clearly a feasible point for the problem (\ref{sa11}). For implementing the first iteration,
we  set $\gamma=f(\phi^{(0)}, v^{B,(0)})/||\phi^{(0)}-v_{RF}(\theta^{(0)})||^2$ ensuring that
the objective
function $f(\phi^{(0)}, v^{B,(0)})$ is of the same magnitude as the penalty term $\gamma ||\phi^{(0)}-v_{RF}(\theta^{(0)})||^2$. As the iterative process continues,  we update $\gamma\rightarrow 1.2\gamma$, whenever $||\phiko-v_{RF}(\thetako)||^2>0.9||\phik-v_{RF}(\thetak)||^2$.}

Fig. \ref{fig:conv_obj_fun_RF4} depicts the convergence pattern of the proposed algorithms in generating Fig. \ref{fig:FC_MR_min_rate_vs_Pref} with $P_{ref} = 38.02$ dBm and $N_{RF} = 4$, while Fig. \ref{fig:conv_penalty_RF4}  depicts
the convergence to zero of the penalty terms. The FC algorithms require more iterations than their AOSA counterparts, because they involve many more decision variables of the phase shifters. {\color{black}Finally, the convergence patterns of objective function (13) in generating Fig. \ref{fig:AOSA_MR_min_rate_vs_RF}, and of objective function (31) in generating Fig. \ref{fig:AOSA_MR_SR_sum_rate_vs_RF}, are depicted in Fig. \ref{fig:conv_obj_fun_no_penalty_RF4}  both for $N_{RF} = 4$ and $P_{ref} = 33.95$ dBm, illustrating the efficacy of the proposed algorithms in resolving these two problems. To offer a concise depiction of the convergence behaviors of the MR and SR, we use the mean rate value for SR maximization.}
	
\section{Conclusions}
A communication network relying on a massive antenna-array at the BS was considered, which supported
multiple users. For energy-efficient delivery of high bit rates over mmWave and sub-Terahertz frequency bands, hybrid beamforming was harnessed, which relied on the concatenation of an array-of-subarrays structured
analog beamformer and a baseband beamformer. The analog beamformer had a low resolution for the sake of a low-complexity practical implementation. To offer a uniform quality-of-service to all users, we  maximized the users' minimum rate. Furthermore, we have shown that our new soft max-min rate based design is computationally attractive, since it is based on scalable algorithms, and it succeeds in attaining an attractive minimum rate and sum rate.
\section*{Appendix: inequality ingredients}
Define the function
\[
\pi(\bx,\bfy)\triangleq \sum_{k=1}^K(1-\frac{||\bx_k||^2}{\bfy_k})
\]
for $\bx\triangleq (\bx_1,\dots, \bx_K)$ and $\bfy\triangleq (\bfy_1,\dots, \bfy_K)$, with $\bx_k\in\mathbb{C}^{N_k}$ and $\bfy_k\in\mathbb{R}$, $k=1,\dots, K$ over the domain
\begin{equation}\label{ap2}
	\{(\bx,\bfy) : ||\bx_k||^2<\bfy_k, k=1,\dots, K\}.
\end{equation}
\begin{myth}\label{th1} In the domain constrained by (\ref{ap2}) the function $\chi(\bx,\bfy)\triangleq
	\ln\pi(\bx,\bfy)$ is concave. Then
	the following inequality holds true for all $(\bx,\bfy)$ and $(\bar{x},\bar{y})$ in the domain constrained by (\ref{ap2}):
	\begin{align}
		\ln\pi(\bx,\bfy)\leq&\ln\pi(\bar{x},\bar{y})
		+\frac{1}{\pi(\bar{x},\bar{y})}\sum_{k=1}^K\frac{||\bar{x}_k||^2}{\bar{y}_k}\nonumber\\
		&+\frac{1}{\pi(\bar{x},\bar{y})}\sum_{k=1}^K\left(-2\frac{\Re\{\bar{x}_k^H\bx_k\}}
		{\bar{y}_k} +\frac{||\bar{x}_k||^2}{\bar{y}_k^2}\bfy_k    \right).\label{ap5}
	\end{align}
\end{myth}
\Prf By \cite[Appendix]{RTKN14}, the function $\pi(\bx,\by)$ is concave. Therefore, $\chi(\bx,\bfy)$ is
concave as the composition of the concave and monotonically increased function $\ln \pi$ and concave function $\pi(\bx,\bfy)$ \cite{Tuybook}. Note that the RHS of (\ref{ap5}) is the linearized function of
the concave function $\ln\pi(\bx,\bfy)$ at $(\bar{x},\bar{y})$, so it is a tight majorant of the LHS \cite{Tuybook}.\qed
For $c>0$ define
\[
\pi_c(\bx,\bfy)=\sum_{k=1}^K(1-\frac{||\bx_k||^2}{||\bx_k||^2+c\bfy_k}).
\]
It follows from (\ref{ap5}) that
the following inequality holds true for all $\bx_k\in\mathbb{C}^{N_k}$, $\bar{x}_k\in\mathbb{C}^{N_k}$,
and $y_k>0$, $\bar{y}_k>0$, $k=1,\dots, K$:
\begin{eqnarray}
	\ln\pi_c(\bx,\bfy)&\leq&
	\ln\pi_c(\bar{x},\bar{y})+\frac{1}{\pi_c(\bar{x},\bar{y})}\sum_{k=1}^K\frac{||\bar{x}_k||^2}
	{c\bar{y}_k+||\bar{x}_k||^2}\nonumber\\
	&&+\frac{1}{\pi_c(\bar{x},\bar{y})}
	\sum_{k=1}^K\left(-2\frac{\Re\{\bar{x}_k^H\bx_k\}}
	{c\bar{y}_k+||\bar{x}_k||^2}\right.\nonumber\\ &&\left.+\frac{||\bar{x}_k||^2}{(c\bar{y}_k+||\bar{x}_k||^2)^2}(c\bfy_k+||\bx_k||^2)\right).\label{ap6}
\end{eqnarray}
Note that the particular case $K=1$ of (\ref{ap6}) is the following inequality, which was derived in \cite{TTN16}:
\begin{eqnarray}\label{inv2}
	\ln\left(1+\frac{||\bx||^2}{\bfy}\right)&\geq& \ln\left(1+\frac{||\bar{x}||^2}{\bar{y}}\right)-\frac{||\bar{x}||^2}{\bar{y}}+
	2\frac{\Re\{\bar{x}^H\bx\}}{\bar{y}}\nonumber\\
	&&-\frac{||\bar{x}||^2}{\bar{y}(||\bar{x}||^2+\bar{y})}(||\bx||^2+\bfy),
\end{eqnarray}
for all $\bx\in\mathbb{C}^N$, $\bfy>0$,  and $\bar{x}\in\mathbb{C}^N$, $\bar{y}>0$.

\bibliographystyle{IEEEtran}
\balance \bibliography{mmwave}
	
\end{document}